\documentclass[11pt]{amsart}
\usepackage{amsbsy,amssymb,amscd,amsfonts,latexsym,amstext,delarray,
amsmath,graphicx} 

\newtheorem{thm}{Theorem}[section]
\newtheorem{prop}[thm]{Proposition}
\newtheorem{cor}[thm]{Corollary}

\numberwithin{equation}{section}

\def\bS{{\mathbb S}}
\def\bT{{\mathbb T}}

\def\C{{\mathbb C}}

\renewcommand{\H}{{\mathbb H}}

\def\Z{{\mathbb Z}}
\def\R{{\mathbb R}}

\def\cA{{\mathcal A}}

\def\cH{{\mathcal H}}

\def\cL{{\mathcal L}}
\def\cM{{\mathcal M}}

\def\cV{{\mathcal V}}

\def\GL{{\rm GL}}

\def\Tr{{\rm Tr}}

\def\elel{(  \downarrow 1)}
\def\dodo{(   \downarrow 3)}
\def\nunu{(   \uparrow 1)}
\def\upup{(   \uparrow 3)}
\def\mass{Y}

\def\fa{{\mathfrak a}}
\def\fb{{\mathfrak b}}
\def\fc{{\mathfrak c}}
\def\fd{{\mathfrak d}}
\def\fe{{\mathfrak e}}

\title{Early universe models from Noncommutative Geometry}
\author{Matilde Marcolli}
\author{Elena Pierpaoli}
\address{Department of Mathematics  \\
California Institute of Technology \\ 
Pasadena, CA 91125, USA}
\email{matilde\@@caltech.edu}
\address{Department of Physics and Astronomy \\
University of Southern California \\
Los Angeles, CA 90089, USA}
\email{pierpaol\@@usc.edu}

\begin{document}
\maketitle

\begin{abstract}We investigate cosmological predictions on the early universe based on the
noncommutative geometry models of gravity coupled to matter. Using the renormalization
group analysis for the Standard Model with right handed neutrinos and Majorana mass
terms, which is the particle physics content of the most recent noncommutative geometry 
models, we analyze the behavior of the coefficients of the gravitational and cosmological
terms in the Lagrangian derived from the asymptotic expansion of the spectral action
functional of noncommutative geometry. We find emergent Hoyle-Narlikar and conformal
gravity at the see-saw scales and a running effective gravitational constant, which affects
the propagation of gravitational waves and the evaporation law of primordial black holes
and provides Linde models of negative gravity in the early universe. The same renormalization
group analysis also governs the running of the effective cosmological constant of the model.
The model also provides a Higgs based slow-roll inflationary mechanism, for which one
can explicitly compute the slow-roll parameters. The particle physics content 
allows for dark matter models based on sterile neutrinos with Majorana mass terms.
\end{abstract}

\tableofcontents

\section{Introduction}

The idea of using noncommutative geometry to give a conceptual mathematical 
formulation of the standard model of elementary particle physics dates back to
the work of Connes \cite{CoSM}. It was shown more recently in \cite{CCM} (see
also \cite{CoSM2}, \cite{Ba}, and Chapter 1 of \cite{CoMa}) 
that the noncommutative geometry model of
particle physics can be made compatible with right handed neutrinos and 
neutrino masses and that the full Lagrangian of the standard model with
Majorana mass terms for the right handed neutrinos can be {\em derived}
by a computation from a very simple input of an almost commutative space,
the product of an ordinary spacetime manifold and a finite noncommutative
space. The resulting physical Lagrangian in this noncommutative geometry
model is obtained from the asymptotic expansion in the energy scale 
$\Lambda$ of a natural action functionals defined on noncommutative
spaces, the {\em spectral action}, \cite{ChCo}.  Among the most interesting features
of these models of particle physics based on noncommutative geometry is the
fact that the physical Lagrangian of the model is completely {\em computed} from
a simple geometric input (the choice of a finite dimensional algebra), so that the
physics is very tightly constrained by the underlying geometry.
For reasons of space, we cannot include here any introductory material
about Noncommutative Geometry, but we suggest the interested readers to look at
the survey paper \cite{CoMa-garden} for a user-friendly introduction based
on examples, as well as the books \cite{Co-book} and \cite{CoMa} for
a more complete treatment. 

We focus here on the noncommutative geometry model obtained in \cite{CCM}.
The corresponding physical Lagrangian computed from the asymptotic expansion
of the spectral action contains the full Standard Model Lagrangian with  additional 
Majorana mass terms for right handed neutrinos, as well as  
gravitational and cosmological terms coupled to matter. The presence of
these terms and their relation to the particle physics content of the model
make this approach of interest to theoretical cosmology. The gravitational terms 
include the Einstein--Hilbert action with a cosmological term, a topological
term related to the Euler characteristic of the spacetime manifold,
and, additionally, a conformal gravity term with the Weyl curvature tensor
and a conformal coupling of the Higgs field to gravity. Without these
last two contributions essentially one would be dealing with the
usual minimal coupling of the standard model to gravity, but the
presence of an additional non-minimal conformal coupling has relevance
to various cosmological models that have been the object of recent
investigation, especially in the context of modified theories of gravity.
Another important way in which this model differs from the minimal coupling
of gravity and matter is the dependence of the coefficients of the gravitational
terms upon the Yukawa parameters of the particle physics content of the model.
This feature, which is our main focus of investigation in the present paper, is 
unique to these noncommutative geometry models and does not have
an analog in other particle physics and cosmology models obtained from  
geometric settings such as string theories, extra dimensions, or brane worlds. 
Some conceptual similarities with these other approaches exist though, in the sense that
in noncommutative geometry models one typically 
modifies ordinary spacetime by taking a product with a noncommutative space
and this may be thought of as another possible way to enrich it with extra dimensions.

The fact that, as shown in \cite{CCM}, the model lives naturally at unification
scale, means that in cosmological terms it provides us with early universe models,
hence it is interesting in terms of possible inflationary mechanisms. Extrapolations
to lower energies are possible using renormalization group analysis, though
extensions to cosmological models of the more recent universe only become
possible when non-perturbative effects in the spectral action are also taken into
account. The main motivation for considering these noncommutative geometry
models in a cosmological context is that the nontrivial dependence of the
cosmological and gravitational parameters on the particle physics content
is, as we mentioned above, significantly different from other physical models,
hence likely to provide inflationary scenarios in the early universe that differ
significantly from other models. 

This paper is the first of a planned series dedicated to an investigation 
of the cosmological implication of the noncommutative geometry models
in particle physics. In the  present paper, we concentrate on a renormalization group analysis
of the coefficients of the gravitational terms in the action. In fact,
the asymptotic formula for the spectral action used in \cite{CCM}
shows that these coefficients are functions of certain parameters,
which in turn depend on the data of the Yukawa parameters of
the standard model with Majorana mass terms for right handed
neutrinos. They also depend on three additional parameters of the 
model, one of which is fixed by a grand unification type condition 
on the coupling constants.

The Yukawa parameters run with the renormalization group equations
of the particle physics model. In particular, since the noncommutative
geometry model lives naturally at unification scale, one can input
boundary conditions at that energy scale and follow the renormalization
group equations towards lower energies and investigate the effect
of this running on the gravitational part of the model. One expects 
that, when running towards lower energies, nonperturbative effects
in the spectral action will progressively become non-negligible. 
This can limit the range of validity of this type of argument based on
the asymptotic expansion alone, and on renormalization group
analysis to cosmological models for the very early universe, that is, 
for sufficiently high energies where the asymptotic expansion holds.  
Any extrapolation to the modern universe would then have to take
into account the full spectral action and not only its asymptotic form.

In the present paper we focus on early universe models and
on the asymptotic form of the spectral action. For the renormalization
analysis, we rely on a detailed study of the renormalization group
equations (RGE) for the extension of the standard model 
with right handed neutrinos and Majorana mass terms
carried out in \cite{AKLRS}, even though their
choice of boundary conditions at unification is different from some
of the boundary conditions assumed in \cite{CCM}. The boundary 
conditions proposed in  \cite{AKLRS} are dictated by particle
physics considerations, while some of the constraints considered
in \cite{CCM} came from analyzing particular geometries, such as
the flat space case. For example, for simplicity the Majorana masses 
were assumed in \cite{CCM} to be degenerate, all of them close to 
unification scale, while here we are mostly interested in the 
non-degenerate case, with three different see-saw scales between 
the electroweak and unification scales, which leads to a more 
interesting behavior of the gravitational terms in the model. 
We plan to return to a more general analysis of the 
RGE flow of \cite{AKLRS} with a wider range of possible boundary 
conditions in followup work. 

The RGE flow of \cite{AKLRS} runs between a unification energy,
taken there to be of the order of $2\times 10^{16}$ GeV, down to
the electroweak scale of $100$ GeV. In terms of cosmological timeline,
we are looking at the behavior of the model between the unification
and the electroweak era. This means that, in terms of matter content,
only the Higgs field and its coupling to gravity is relevant, so we 
mostly concentrate on the part of the Lagrangian of \cite{CCM} that 
consists of these terms.  Since this era of the
early universe is believed to include the inflationary epoch, we
look especially at different possible inflationary scenarios provided by this 
noncommutative geometry model. 

Our main results in this first paper are to show that, using the information on
the particle physics content, it is possible to obtain cosmological models of
the early universe with variable gravitational and cosmological constant,
hence providing a range of different mechanisms for inflation, realized 
by the running of the effective gravitational constant and by its
coupling to the Higgs field, or by the running of the effective cosmological constant
of the model, or by a combination of these. We also show phenomena where, near 
particular energy scales and for special geometries, the usual
Einstein--Hilbert action ceases to be the dominant contribution and the model 
comes to be dominated, at certain scales, by conformal gravity and
an emergent Hoyle--Narlikar cosmology. 
We discuss how the running of the gravitational parameters of the
model influences the behavior of the evaporation of primordial black holes 
by Hawking radiation. While the type of effects that we see in this model,
which depend on the presence of variable effective gravitational and
cosmological constants, are qualitatively similar to scenarios of negative
gravity in the early universe previously analyzed in theoretical cosmology
(\cite{Barrow}, \cite{BeKoKo}, \cite{Carr}, \cite{dSHW}, \cite{Dolgov}, \cite{EFP}, 
\cite{Linde}, \cite{Loh}, \cite{Mann}, \cite{Novikov}, \cite{OvCo}, \cite{Pollock}, 
\cite{SaLo}, \cite{Zee}), 
the mechanism that produces these effects in the noncommutative geometry 
model is substantially different from those described in these earlier references, 
which makes the quantitative behavior also different and distinguishable from
other models.  In fact, most of the effects we investigate in this paper
depend directly on the expression of the coefficients of the gravitational
and bosonic terms in the asymptotic expansion of the spectral action 
in terms of the Yukawa parameters of the underlying particle physics model.
This is a {\em purely geometric} property of this model and it comes directly
from the presence of the ``small extra dimensions" in the form of the zero
dimensional (but K-theoretically six dimensional) finite noncommutative space
in addition to the extended spacetime dimensions. 

While the energy range where the renormalization group analysis applies limits
the results based only on the perturbative expansion of the spectral action to
early universe models, if some of the results obtained in this paper persist when
non-perturbative effects in the spectral action become significant, they may provide 
possible dark energy and dark matter predictions. For instance, the behavior of
the variable effective cosmological constant may lead to dark energy scenarios 
in the more recent universe. Moreover, we show that the
particle physics content of the  model is consistent with dark matter models based
on right handed neutrinos with Majorana mass terms in \cite{Kusenko},
\cite{Shapo}, \cite{ShapoTka}. In fact, the particle content is the same as in the
$\nu$MSM model with three active and three sterile neutrinos. What is needed
in order to relate the model of \cite{CCM} to these dark matter models is a choice
of boundary conditions that makes it possible for at least one of the Majorana 
masses to descent to somewhere near the electroweak scale, hence providing
sterile neutrinos with the characteristics required to give acceptable dark matter
candidates (see the analysis in \cite{Kusenko}). We will return to a closer
analysis of dark energy and dark matter implications of the noncommutative
geometry models in a planned continuation of this work.

In a related but different direction, recent work on some cosmological aspects 
of the model of \cite{CCM} was also done in \cite{NeSa}. 

\smallskip

{\bf Acknowledgment.} Part of this work was carried out during visits of the
first author at the Mathematical Sciences Research Institute in Berkeley
and at the Max Planck Institute for Mathematics in Bonn. The hospitality 
and support of both institutions is gratefully acknowledged. 

\section{The asymptotic formula for the spectral action and the gravitational parameters}

We recall here briefly the main aspects of the noncommutative geometry model of
particle physics derived in \cite{CCM} that we need to use in the rest of the paper.
We refer the reader to \cite{CCM} and to Chapter 1 of \cite{CoMa} for a detailed
treatment. The reader who wishes to skip this preliminary part can start directly with
the asymptotic expansion of the spectral action recalled in \S \ref{asymptSec}, which is
what we concentrate on in the rest of the paper, but  we prefer to add a few words on the
derivation of the model via noncommutative geometry for the sake of completeness.

\subsection{Spectral triples and the spectral action functional}

The particle physics models based on noncommutative geometry, both the
original one of \cite{CoSM} and the new one of \cite{CCM} that incorporates
right handed neutrinos and neutrino mixing with Majorana mass terms, are 
based on the formalism of {\em spectral triples}. These were introduced by
Connes \cite{CoS3} as an extension of the notion of Riemannian manifold
to noncommutative geometry. The data $({\mathcal A},{\mathcal H},D)$
defining a (real) spectral triple are summarized as follows.

\begin{itemize}
\item $\cA$ is an involutive algebra with unit. Requiring the algebra to be
unital corresponds to working with compact manifolds. (Extensions of
the notion of spectral triple to non-unital cases have also been developed.)
\item $\cH$ is a separable Hilbert space endowed with a representation 
$\pi:\cA \to \cL(\cH)$ of the algebra $\cA$ by bounded linear operators.
\item $D=D^\dag$ is a linear self-adjoint operator acting on $\cH$. Except for finite dimensional cases, 
$D$ is in general not a bounded operator, hence it is only defined on a dense domain.
\item $D$ has compact resolvent: $(1+D^2)^{-1/2}$ is a compact operator.
\item The commutators $[\pi(a),D]$ are bounded operators for all $a\in \cA$.
\item The spectral triple is even if there is on $\cH$ a $\Z/2$-grading $\gamma$ 
satisfying $[\gamma,\pi(a)]=0$ and $D\gamma =-\gamma D$.
\item The spectral triple has a real structure if there is an antilinear isomorphism 
$J: \cH \to \cH$ with $J^2 = \varepsilon$, $JD = \varepsilon' DJ$, and 
$J\gamma = \varepsilon'' \gamma J$, where the signs $\epsilon$, $\epsilon'$, and $\epsilon''$
determine the KO-dimension modulo $8$ of the spectral triple, according to by the table
\begin{center}
\begin{tabular}
{|c| r r r r r r r r|} \hline {\bf n }&0 &1 &2 &3 &4 &5 &6 &7 \\
\hline \hline
$\varepsilon$  &1 & 1&-1&-1&-1&-1& 1&1 \\
$\varepsilon'$ &1 &-1&1 &1 &1 &-1& 1&1 \\
$\varepsilon''$&1 &{}&-1&{}&1 &{}&-1&{} \\  \hline
\end{tabular}
\end{center}
\item The Hilbert space $\cH$ has an $\cA$-bimodule structure with respect to the action of $\cA$
defined by $b^0 = J b^* J^{-1}$ and satisfying the commutation condition $[a,b^0] = 0$ for all $a$ and 
$b$ in $\cA$.
\item The operator $D$ satisfies the order one condition $[[D,a],b^0] = 0$, for all $a,b\in \cA$.
\end{itemize}

Commutative geometries, which in this context means ordinary Riemannian manifolds,
can be described as spectral triples: for a compact spin Riemannian
manifold $X$ the associated spectral triple $(C^\infty(X), L^2(X,S), D_X)$ is given by 
the algebra of smooth functions, the Hilbert space of square integrable spinors, and the
Dirac operator. The metric tensor can be recovered from these data. For an even
dimensional manifold $\gamma_X=\gamma_5$ is the grading $\cH=\cH^+\oplus \cH^-$
on the spinor bundle given by the usual chirality operator $\gamma_5$ and the real
structure $J_X$ is the charge conjugation operator. Examples of
spectral triples associated to objects that are not manifolds include a wide range of geometries
such as quantum groups, fractals, or noncommutative tori. As we recall in \S \ref{ncgSec}
below, the spectral triples involved in the particle physics models are of a very special form
which is {\em almost commutative}, namely a product of an ordinary manifold with a
small noncommutative space.

\smallskip

It was shown by Chamseddine and Connes \cite{ChCo} that there is a natural action functional
on a spectral triple. This {\em spectral action} functional is defined as
$\Tr(f(D/\Lambda))$, where $f>0$ is a cutoff function and $\Lambda$ is the energy scale.
There is an asymptotic formula for the spectral action, for large energy $\Lambda$, of the form
\begin{equation}\label{SpActLambda}
\Tr(f(D/\Lambda))\sim \sum_{k\in {\rm DimSp}} f_{k} \Lambda^k {\int\!\!\!\!\!\!-} |D|^{-k} + f(0) \zeta_D(0)+ o(1),
\end{equation}
where $f_k= \int_0^\infty f(v) v^{k-1} dv$ are the momenta of the function $f$ and the
noncommutative integration is defined in terms of residues of zeta functions 
\begin{equation}\label{zetaD}
 \zeta_{a,D} (s) = \Tr (a \, |D|^{-s}). 
\end{equation}
The sum in \eqref{SpActLambda} is over points in the {\em dimension spectrum} of the
spectral triple, which is a refined notion of dimension for noncommutative spaces, consisting
of the set of poles of the zeta functions \eqref{zetaD}.

\subsection{The noncommutative space of the model}\label{ncgSec}

The main result of \cite{CCM} is a complete derivation of the full standard
model Lagrangian with additional right handed neutrino, lepton mixing matrix
and Majorana mass terms, by a {\em calculation} starting from a very simple
geometric input. The initial {\em ansatz} used in \cite{CCM} is the choice of
a finite dimensional algebra, the left-right symmetric algebra
\begin{equation}\label{LRalg}
\cA_{LR}= \C \oplus \H_L \oplus \H_R \oplus M_3(\C),
\end{equation}
where $\H_L$ and $\H_R$ are two copies of the real algebra of quaternions.

The representation is then naturally determined by taking the sum $\cM$ of all
the inequivalent irreducible odd spin representations of $\cA_{LR}$, so that
the only further input that one needs to specify is the number $N$ of generations.
The (finite dimensional) Hilbert space is then given by $N$ copies of $\cM$,
$$ \cH_F = \oplus^N \cM . $$
The Hilbert space with the $\cA_{LR}$ action splits as a sum $\cH_F=\cH_f \oplus \cH_{\bar f}$
of matter and antimatter sectors, and an 
orthogonal basis of $\cH_f$ gives all the fermions of the particle
physics model
\begin{equation}\label{fermions}
\begin{array}{ll}
\nu_L = | \uparrow \rangle_L \otimes {\bf 1}^0 & \nu_R =| \uparrow \rangle_R 
\otimes {\bf 1}^0 \\
e_L = | \downarrow \rangle_L \otimes {\bf 1}^0 & e_R =| \downarrow \rangle_R \otimes {\bf 1}^0 \\
u_L =| \uparrow\rangle_L \otimes {\bf 3}^0 & u_R =| \uparrow \rangle_R \otimes {\bf 3}^0 \\
d_L =| \downarrow\rangle_L \otimes {\bf 3}^0 & d_R =| \downarrow \rangle_R \otimes {\bf 3}^0,
\end{array}
\end{equation}
respectively giving the neutrinos, the charged leptons, the u/c/t quarks, and the d/s/b quarks in
terms of the representation of $\cA_{LR}$.  Here $|\uparrow\rangle$ and $|\downarrow\rangle$
are the basis of the ${\bf 2}$ representation of $\H$ where the action of $\lambda\in 
\C \subset \H$ is,  respectively, by $\lambda$ or $\bar\lambda$, and ${\bf 1}^0$ and
${\bf 3}^0$ are the actions of $\C$ and $M_3(\C)$, respectively, through the 
representation $a^0=J a^* J^{-1}$.

The $\Z/2$-grading  $\gamma_F$ exchanges the left and right chirality of fermions and
the real structure operator $J_F$ exchanges the matter and antimatter sectors and performs
a complex conjugation. These properties of $\gamma_F$ and $J_F$ suffice to determine
the KO-dimension modulo $8$ of the resulting spectral triple and an interesting aspect is that,
unlike in the earlier particle physics models based on noncommutative geometry, in this
case the KO-dimension is $6$ modulo $8$, although the metric dimension is zero.

The order one condition on the Dirac operator is seen in \cite{CCM} as a coupled
equation for a subalgebra $\cA_F \subset \cA_{LR}$ and a Dirac operator and it
is shown that there is a unique subalgebra of maximal dimension that allows for
the order one condition to be satisfied. The algebra $\cA_F$ is of the form
\begin{equation}\label{AF}
\cA_F = \C \oplus \H \oplus M_3(\C),
\end{equation}
where the first summand embeds diagonally into $\C \oplus \H$ in $\cA_{LR}$,
thus breaking the left-right symmetry. It is expected, though presently not known,
that this symmetry breaking should be dynamical. This geometric argument identifying
the maximal algebra on which the order one condition can be satisfied was later extended
in \cite{ChCo2} to more general {\em ansatz} algebras than $\cA_{LR}$, but with the 
same resulting $\cA_F$. 

\subsection{Dirac operators: Yukawa parameters and Majorana masses}

The selection of the subalgebra $\cA_F$ for the order one condition for the
Dirac operator is what produces geometrically in this model the Majorana mass
terms for right handed neutrinos. In fact, one has in \cite{CCM} a complete
classification of the possible Dirac operators on the noncommutative space
$(\cA_F,\cH)$ compatible with $\gamma_F$ and $J_F$ (see also \cite{Cacic}
for a more general discussion of moduli spaces of Dirac operators for
finite spectral triples). These are all of the form
$$ D(\mass)=\left(\begin{matrix}\bS &\bT^\dag\\
\bT &\bar \bS\end{matrix} \right), $$
with $\bS = \bS_1 \, \oplus (\bS_3 \otimes 1_3)$ and  $\bT=Y_R : |\nu_R \rangle \to J_F\,|\nu_R \rangle$,
and with $\bS_1$ and $\bS_3$ respectively of the form
 $$ \bS_1 = \left(\begin{matrix}0&0&\mass^\dag_{\nunu}&0\\
0&0&0&\mass^\dag_{\elel}\\
\mass_{\nunu}&0&0&0\\
0&\mass_{\elel}&0&0&\end{matrix} \right) $$
$$ \bS_3 = \left(\begin{matrix}0&0&\mass^\dag_{\upup}&0\\
0&0&0&\mass^\dag_{\dodo}\\
\mass_{\upup}&0&0&0\\
0&\mass_{\dodo}&0&0&\end{matrix} \right) . $$
Here the $N\times N$-matrices involved in the expression of $\bS_1$ and $\bS_3$
are the Yukawa matrices that give Dirac masses and mixing angles.
These are matrices in $\GL_3(\C)$ in the case of $N=3$ generations:
$\mass_{e}=\mass_{\elel}$ is the Yukawa matrix for the charged leptons,
$\mass_{\nu}=\mass_{\nunu}$ for the neutrinos, 
$\mass_{d}=\mass_{\dodo}$ for the d/s/b quarks, and 
$\mass_{u}=\mass_{\upup}$ for the u/c/t quarks. Moreover, the remaining term 
$M=Y_R^T$, with $T$ denoting transposition, gives the matrix $\bT$ 
in $D(\mass)$ and is the symmetric matrix of the Majorana mass terms for right 
handed neutrinos. 

\medskip

Thus, the model of \cite{CCM} has three active and three sterile neutrinos as 
in the $\nu$MSM model, see \cite{Kusenko}, \cite{Shapo}, \cite{ShapoTka}, though
in \cite{CCM}, unlike in the $\nu$MSM model, it is assumed that the three sterile 
neutrinos all have masses well above the electroweak scale. 
The see-saw relation $Y_\nu^T M^{-1} Y_\nu$  for neutrino masses
is obtained in \cite{CCM} geometrically from the fact that the restriction
of the Dirac operator $D(\mass)$ to the subspace of $\cH_F$ spanned by $\nu_R$,
$\nu_L$, $\bar \nu_R$, $\bar \nu_L$ is of the form
\begin{equation}\label{seesawCCM}
\left( \begin{array}{cccc}
0 & M_{\nu}^\dag & \bar M_R^\dag & 0 \\
M_{\nu} & 0 & 0 & 0 \\
\bar M_R & 0 & 0 & \bar M_{\nu}^\dag \\
0 & 0 & \bar M_\nu & 0 
\end{array}\right),
\end{equation}
where $M_{\nu}$ is the neutrino mass matrix, see Lemma 1.225 of \cite{CoMa}.
We return to discuss the relation of the model of \cite{CCM} to the $\nu$MSM model
of \cite{Shapo}, \cite{ShapoTka} and to other sterile neutrinos scenarios of \cite{Kusenko}
in the context of dark matter models in cosmology, see \S \ref{DarkSec} below.

\medskip

The spectral triple that determines the physical Lagrangian of the model through
the asymptotic expansion of the spectral action is then the {\em product geometry}
$X\times F$, of a 4-dimensional spacetime $X$, identified with the spectral triple
$(C^\infty(X), L^2(X,S), D_X)$, and the finite noncommutative
space $F$ defined by the spectral triple $(\cA_F,\cH_F,D_F)$ with $D_F$ of the 
form $D(\mass)$ as above. The product is given by the cup product spectral triple
$(\cA,\cH,D)$ with sign $\gamma$ and real structure $J$
\begin{itemize}
\item $\cA=C^{\infty} (X) \otimes \cA_F = C^{\infty} (X,\cA_F)$
\item $\cH = L^2 (X,S) \otimes \cH_F = L^2 (X,S\otimes \cH_F)$
\item $D = D_X \otimes 1 + \gamma_5 \otimes D_F$
\item $J=J_X \otimes J_F$ and $\gamma = \gamma_5 \otimes \gamma_F$.
\end{itemize}

\smallskip

The action functional considered in \cite{CCM} to obtain the physical Lagrangian
has a bosonic and a fermionic part, where the bosonic part is given by the spectral
action functional with inner fluctuations of the Dirac operator and the fermionic
part by the pairing of the Dirac operator with fermions, 
\begin{equation}\label{ActionCCM}
\Tr(f(D_A/\Lambda))  + \frac 12 \langle J \tilde\xi,D_A \tilde\xi\rangle.
\end{equation}
Here $D_A=D+A+\varepsilon' \,J\,A\,J^{-1}$ is the Dirac operator with inner fluctuations given
by the gauge potentials of the form $A=A^\dag=\sum_k a_k[D,b_k]$, for elements $a_k,b_k\in \cA$.
The fermionic term $\langle J \tilde\xi,D_A \tilde\xi\rangle$ should be seen as a pairing of
classical fields $\tilde\xi \in \cH^+=\{ \xi \in \cH\,|\, \gamma \xi =\xi\}$, 
viewed as Grassman variables. This is a common way of treating Majorana spinors via Pfaffians,
see  \S 16.2 of \cite{CoMa}.

\medskip

While this fermionic part is very important for the particle physics content of the model,
as it delivers all the fermionic terms in the Lagrangian of the Standard Model, 
for our purposes related to cosmological models of the early universe, it will 
suffice of concentrate only on the bosonic part of the action, given by the spectral action
term $\Tr(f(D_A/\Lambda))$, since during a good part of the cosmological period between 
the unification and the electroweak epoch the Higgs field is the matter content that will 
be mostly of relevance, \cite{Guth}.

\subsection{Parameters of the model}\label{paramSec}

As we have recalled above, the geometric parameters describing the possible
choices of Dirac operators on the finite noncommutative space $F$ correspond
to the Yukawa parameters of the particle physics model and the Majorana mass terms
for the right handed neutrinos. We recall here some
expressions of these parameters that appear in the asymptotic expansion of
the spectral action and that we are going to analyze more in detail later in the paper.
We define functions $\fa$, $\fb$, $\fc$, $\fd$, $\fe$ of the matrices $\mass_u$, $\mass_d$,
$\mass_{\nu}$, $\mass_e$ and of the Majorana masses $M$ in the following way:
\begin{equation}\label{abcde}
\begin{array}{rl}
  \fa =& \,\Tr(\mass_{\nu}^\dag \mass_{\nu}+\mass_{e}^\dag \mass_{e}
  +3(\mass_{u}^\dag \mass_{u}+\mass_{d}^\dag \mass_{d})) \\[2mm]
  \fb =& \,\Tr((\mass_{\nu}^\dag \mass_{\nu})^2+(\mass_{e}^\dag \mass_{e})^2+3(\mass_{u}^\dag \mass_{u})^2+3(\mass_{d}^\dag \mass_{d})^2) \\[2mm]
  \fc =& \Tr(M M^\dag)  \\[2mm]
  \fd =& \Tr((M M^\dag)^2)  \\[2mm]
  \fe =& \Tr(M M^\dag \mass_{\nu}^\dag \mass_{\nu}) .
\end{array}
\end{equation} 
In addition to these parameters, whose role we describe in \S \ref{asymptSec},
we see clearly from \eqref{SpActLambda} that the asymptotic formula for the 
spectral action depends on parameters $f_k$ given by the momenta of the
cutoff function $f$ in the spectral action. Since the noncommutative space here
is of the simple form $X\times F$, the only contributions to the dimension
spectrum, hence to the asymptotic formula for the spectral action come
from three parameters $f_0$, $f_2$, $f_4$, where $f_0=f(0)$ and for $k>0$
$$ f_k =\int_0^\infty f(v) v^{k-1} dv . $$

\subsection{The asymptotic expansion of the spectral action}\label{asymptSec}

It was proved in \cite{CCM} that the asymptotic formula \eqref{SpActLambda} applied
to the action functional $\Tr(f(D_A/\Lambda))$ of the product geometry $X\times F$ gives
a Lagrangian of the form
\begin{equation}\label{asymptSA}
\begin{array}{rl}
S =& \displaystyle{ \frac{1}{\pi^2}(48\,f_4\,\Lambda^4-f_2\,\Lambda^2\,\fc+\frac{
f_0}{4}\,\fd) }\,\int \,\sqrt g\,d^4 x \\[3mm]
    +& \, \displaystyle{
     \frac{96\,f_2\,\Lambda^2 -f_0\,\fc}{ 24\pi^2} }\, \int\,R
 \, \sqrt g \,d^4 x  \\[3mm]
    +& \, \displaystyle{
    \frac{f_0 }{ 10\,\pi^2} } \int\,(\frac{11}{6}\,R^* R^* -3 \, C_{\mu
\nu \rho \sigma} \, C^{\mu \nu \rho \sigma})\, \sqrt g \,d^4 x \\[3mm]
 +&  \, \displaystyle{
 \frac{(- 2\,\fa\,f_2
  \,\Lambda^2\,+ \,\fe\,f_0 )}{ \pi^2} } \int\,  |\varphi|^2\, \sqrt g \,d^4 x 
\\[3mm]
    +&  \, \displaystyle{
    \frac{f_0 \fa}{ 2\,\pi^2} } \int\,  |D_{\mu} \varphi|^2\, 
\sqrt g \,d^4 x \\[3mm]
   -&  \displaystyle{
   \frac{f_0 \fa}{ 12\,\pi^2} } \int\, R \, |\varphi|^2 \, \sqrt g \,d^4 x
 \\[3mm]
    +&\, \displaystyle{ \frac{f_0 \fb}{ 2\,\pi^2} } \int |\varphi|^4 \, \sqrt g \,d^4 x \\[3mm]
+& \, \displaystyle{ \frac{f_0 }{ 2\,\pi^2} } \int\,(g_{3}^2 \, G_{\mu \nu}^i \, 
G^{\mu \nu i} +  g_{2}^2 \, F_{\mu
\nu}^{ \alpha} \, F^{\mu \nu  \alpha}+\, \frac{5}{ 3} \,
g_{1}^2 \,  B_{\mu \nu} \, B^{\mu \nu})\, \sqrt g \,d^4 x ,
\end{array}
\end{equation}
We see from this expansion how the coefficients of all the terms in this
resulting action functional depend on the Yukawa and Majorana parameters 
through their combinations of the form $\fa$, $\fb$, $\fc$, $\fd$, $\fe$ defined 
as in \eqref{abcde}, and from the three additional parameters $f_0$, $f_2$, $f_4$.

The term of \eqref{asymptSA} with the Yang--Mills action for 
the gauge bosons,
$$ \frac{f_0 }{ 2\,\pi^2} \int\,(g_{3}^2 \, G_{\mu \nu}^i \, 
G^{\mu \nu i} +  g_{2}^2 \, F_{\mu
\nu}^{ \alpha} \, F^{\mu \nu  \alpha}+\, \frac{5}{ 3} \,
g_{1}^2 \,  B_{\mu \nu} \, B^{\mu \nu})\, \sqrt g \,d^4 x, $$
contains  the coupling constants 
$g_1$, $g_2$, $g_3$ of the three forces.  
As shown in \cite{CCM}, the standard normalization
of these Yang--Mills terms gives the GUT relation
between the three coupling constants and fixes the fact that
this model lives naturally at a {\em preferred energy scale} given by the
unification scale $\Lambda=\Lambda_{unif}$.
The normalization of the Yang--Mills terms fixes the value of the parameter $f_0$ to
depend on the common value $g$ at unification of the coupling constants: as shown in
\S 4.5 and \S 5.1 of \cite{CCM} one obtains
\begin{equation}\label{unification}
 \frac{g^2 f_0}{2\pi^2} = \frac{1}{4}. 
\end{equation}
One also normalizes the kinetic term for the Higgs as in \cite{CCM} by the
change of variables $H = \frac{\sqrt{\fa f_0}}{\pi} \varphi$ 
to get $\frac{1}{2} \int |D H|^2 \sqrt{g} d^4x$. 

The normalization of the Yang--Mills terms and of the kinetic term of the Higgs
then gives, at unification scale, an action functional of the form
\begin{equation}\label{SAvarchange}
\begin{array}{rl}
S =& \displaystyle{\frac{1}{2\kappa_0^2}  \int\,R
 \, \sqrt g \,d^4 x + \gamma_0 \,\int \,\sqrt g\,d^4 x } \\[3mm]
    +& \displaystyle{ \alpha_0 \int C_{\mu
\nu \rho \sigma} \, C^{\mu \nu \rho \sigma} \sqrt g \,d^4 x + \tau_0 \int R^* R^* \sqrt g \,d^4 x } \\[3mm]
 +& \displaystyle{ \frac{1}{2} \int\,  |D H|^2\, 
\sqrt g \,d^4 x -  \mu_0^2 \int\,  |H|^2\, \sqrt g \,d^4 x }
\\[3mm]
  -&  \displaystyle{ \xi_0 \int\, R \, |H|^2 \, \sqrt g \,d^4 x
+ \lambda_0  \int |H|^4 \, \sqrt g \,d^4 x } \\[3mm]
+& \displaystyle{ \frac{1}{4} \int\,(G_{\mu \nu}^i \, 
G^{\mu \nu i} +  F_{\mu
\nu}^{ \alpha} \, F^{\mu \nu  \alpha}+\, B_{\mu \nu} \, B^{\mu \nu})\, \sqrt g \,d^4 x },
\end{array}
\end{equation}
where the coefficients are now 
\begin{equation}\label{coeffsrun}
\begin{array}{ll}
\frac{1}{2\kappa_0^2} = & \displaystyle{\frac{96 f_2 \Lambda^2 - f_0 \fc}{24\pi^2}} \\[3mm]
\gamma_0 = & \displaystyle{ \frac{1}{\pi^2}(48 f_4 \Lambda^4 - f_2 \Lambda^2 \fc 
+\frac{f_0}{4} \fd) } \\[3mm]
\alpha_0 = & \displaystyle{ - \frac{ 3 f_0}{10\pi^2} } \\[3mm]
\tau_0 =& \displaystyle{\frac{11 f_0}{60 \pi^2}} \\[3mm]
\mu_0^2 = &  \displaystyle{ 2 \frac{f_2 \Lambda^2}{f_0} - \frac{\fe}{\fa} } \\[3mm]
\xi_0 = & \frac{1}{12} \\[3mm]
\lambda_0 = & \displaystyle{ \frac{\pi^2 \fb}{2 f_0 \fa^2} } ,
\end{array}
\end{equation}
again as a function of the Yukawa and Majorana parameters through the
coefficients $\fa$, $\fb$, $\fc$, $\fd$, $\fe$ of  \eqref{abcde}, and of the two 
remaining free parameters of the model, $f_2$ and $f_4$, after the value
of $f_0$ has been fixed by the unification condition.

\section{Renormalization group and running parameters}\label{RGEflowSec}

All the Yukawa parameters $Y_u$, $Y_d$, $Y_{\nu}$, $Y_e$, as well as the
Majorana mass terms $M$ are subject to running with the renormalization group
equations (RGE) dictated by the particle physics content of the model, in this case
the Standard Model with additional right handed neutrinos with Majorana mass 
terms. Consequently, also the parameters $\fa$, $\fb$, $\fc$, $\fd$, and $\fe$ of
\eqref{abcde} run with the renormalization group flow as functions of $\Lambda$,
with assigned initial conditions at $\Lambda=\Lambda_{unif}$, which is the
preferential energy scale of the model. 

Some estimates based on renormalization group analysis were obtained already
in \cite{CCM}, for the Higgs and the top quark masses, but those were based, in
first approximation, on just the renormalization group equations at 1-loop for the minimal
Standard Model.

In this section we analyze the running of the parameters of the model with the
renormalization group flow, using the full RGE of the extension of the Standard 
Model by right handed neutrinos and Majorana masses. There is an extensive 
literature available in particle physics on the relevant RGE analysis, see for
instance \cite{AKLR}, \cite{BLP}, \cite{CCIQ}. We use here a more detailed 
analysis of the renormalization group flow, again to one-loop order, for the 
Standard Model with additional Majorana mass terms for right handed 
neutrinos, as given in \cite{AKLRS} and implemented by the authors of
\cite{AKLRS} in the Mathematica package {\tt http://www.ph.tum.de/\char126 rge/REAP/}.

The full renormalization group equations for this particle physics model 
have beta functions given by
\begin{equation}\label{RGEgi}
 16 \pi^2 \, \,  \beta_{ g_i } = b_i \, g_i^3  \ \ \  \text{ with }
(b_{SU(3)}, b_{SU(2)}, b_{U(1)}) = ( -7, - \frac{19}{6}, \frac{41}{10}), 
\end{equation}
where \cite{AKLRS} is using here a different normalization from \cite{CCM}
and the factor $5/3$ has been now included in $g_1^2$. Thus, as for
the minimal Standard Model, at 1-loop order the RGE for the coupling
constants uncouple from those of the other parameters. We then have for
the Yukawa matrices
\begin{equation}\label{RGEYu}
 16 \pi^2 \, \,  \beta_{\mass_u} =  
\mass_u(\frac{3}{2} \mass_u^\dag \mass_u - \frac{3}{2} \mass_d^\dag 
\mass_d + \fa - 
    \frac{17}{20} g_1^2 - \frac{9}{4} g_2^2 - 8g_3^2 ) 
\end{equation}
\begin{equation}\label{RGEYd}    
16 \pi^2 \, \,  \beta_{\mass_d} = 
\mass_d (\frac{3}{2} \mass_d^\dag \mass_d - \frac{3}{2} \mass_u^\dag 
\mass_u +  \fa -
\frac{1}{4}g_1^2 - \frac{9}{4}g_2^2 - 8 g_3^2 ) 
\end{equation}
\begin{equation}\label{RGEYnu}
 16 \pi^2 \, \,  \beta_{\mass_{\nu}} =   \mass_{\nu} (
\frac{3}{2}\mass_{\nu}^\dag \mass_{\nu}- \frac{3}{2}
\mass_e^\dag \mass_e + \fa - \frac{9}{20} g_1^2 - \frac{9}{4} g_2^2 ) 
\end{equation}
\begin{equation}\label{RGEYe}
 16 \pi^2 \, \,  \beta_{\mass_e} = \mass_e (
\frac{3}{2}\mass_e^\dag \mass_e- \frac{3}{2}
\mass_{\nu}^\dag \mass_{\nu}  +\fa  -\frac{9}{4} g_1^2 - \frac{9}{4} g_2^2) .
\end{equation}
The RGE for the Majorana mass terms has beta function
\begin{equation}\label{RGEM}
 16 \pi^2 \, \,  \beta_{M} = 
\mass_\nu \mass_\nu^\dag M + M (\mass_\nu \mass_\nu^\dag)^T 
\end{equation}
and the one for the Higgs self coupling $\lambda$ is given by
\begin{equation}\label{RGElambda}
 16 \pi^2 \, \,  \beta_{\lambda} = 6 \lambda^2 - 3\lambda (3 g_2^2 + \frac{3}{5} g_1^2) + 
 3 g_2^4 + \frac{3}{2} (\frac{3}{5} g_1^2 + g_2^2)^2 + 4\lambda \fa -  8 \fb 
\end{equation}

In the treatment of the renormalization group analysis given in the references mentioned above, 
one assumes that the Majorana mass terms are non-degenerate, which means that there 
are different see-saw scales at decreasing energies in between unification and the electroweak 
scale. In between these see-saw scales, one considers different {\em effective field 
theories}, where the heaviest right handed neutrinos are integrated out when one passes
below the corresponding see-saw scale. In practice the procedure for computing the
RGE for this type of particle physics models can be summarized as follows.
\begin{itemize}
\item Run the renormalization group flow down from unification energy $\Lambda_{unif}$ to 
first see-saw scale determined by the largest eigenvalue of $M$, using assigned boundary
conditions at unification.
\item Introduce an effective field theory where $Y_\nu^{(3)}$ is obtained by removing the
last row of $Y_\nu$ in the basis where $M$ is diagonal and $M^{(3)}$ is obtained by removing the 
last row and column.
\item Restart the induced renormalization group flow given by the equations 
\eqref{RGEYu}--\eqref{RGElambda} with $Y_\nu$ replaced by $Y_\nu^{(3)}$ and
$M$ replaced by $M^{(3)}$ and with matching boundary conditions at the first
see-saw scale. Run this renormalization group flow down to second see-saw scale.
\item Introduce a new effective field theory with $Y_\nu^{(2)}$ and $M^{(2)}$ obtained by
repeating the above procedure starting from $Y_\nu^{(3)}$ and $M^{(3)}$. 
\item Run the induced renormalization group flow for these fields with matching
boundary conditions at the second see-saw scale, down until the first and lowest see-saw scale.
\item Introduce again a new effective field theory with $Y_\nu^{(1)}$ and $M^{(1)}$ and 
matching boundary conditions at the first see saw scale.
\item Run the induced RGE down to the electroweak energy $\Lambda=\Lambda_{ew}$.
\end{itemize}
The procedure illustrated here assumes that the three see-saw scales are all located
between the unification and the electroweak scale, that is, that all the sterile neutrinos
are heavy. 

Notice that there is a difference between the boundary conditions assumed in
\cite{AKLRS} for $M$ at unification energy and the assumption made in \cite{CCM} 
on the Majorana mass terms. In fact, in \S 5.5 of  \cite{CCM} under the assumptions 
of flat space and with Higgs term $|H|^2$ sufficiently small, it is shown that one can
estimate from the equations of motion of the spectral action that the largest Majorana mass 
can be as high as the unification energy, while in \cite{AKLRS} the unification scale
is taken at around $10^{16}$ GeV but the top see-saw scale is around $10^{14}$ GeV.
For the purpose of the present paper we only work with the boundary conditions 
of \cite{AKLRS}, while a more extensive study of the same RGE with a broader range 
of boundary conditions will be considered elsewhere. We report the explicit boundary
conditions of \cite{AKLRS} at $\Lambda=\Lambda_{unif}$ in the appendix.

\subsection{Running parameters and see-saw scales}\label{runparamSec}

In the following, we assume as in \cite{CCM} that the value of the parameter $f_0$ is
fixed by the relation \eqref{unification}. In terms of the boundary conditions 
at unification used in \cite{AKLRS}, this gives as value for $f_0$ either
one of the following
\begin{equation}\label{f03ways}
f_0 =\frac{\pi^2}{2 g_1^2}= 8.52603, \ \ \
f_0 = \frac{\pi^2}{2 g_2^2}= 9.46314, \ \ \  f_0 = \frac{\pi^2}{2 g_3^2}=9.36566.
\end{equation}
These three different choices come from the fact that, as it is well known, the values
for the three coupling constants do not exactly meet in the minimal Standard
Model, nor in its variant with right handed neutrinos and Majorana mass terms. 
Notice that here we are already including the factor of $5/3$ in $g_1^2$,
unlike in \cite{CCM}, where we have $g_3^2 = g_2^2 = \frac{5}{3} g_1^2$ at unification.
We shall perform most of our explicit calculations in the following using the
first value for $f_0$ and it is easy to check that replacing it with either one of the
others will not affect significantly any of the results. The parameters $f_2$ and $f_4$
remain free parameters in the model and we discuss in \S \ref{CosmoSec} how they
can be varied so as to obtain different possible cosmological models.

\begin{figure}
\includegraphics[scale=0.5]{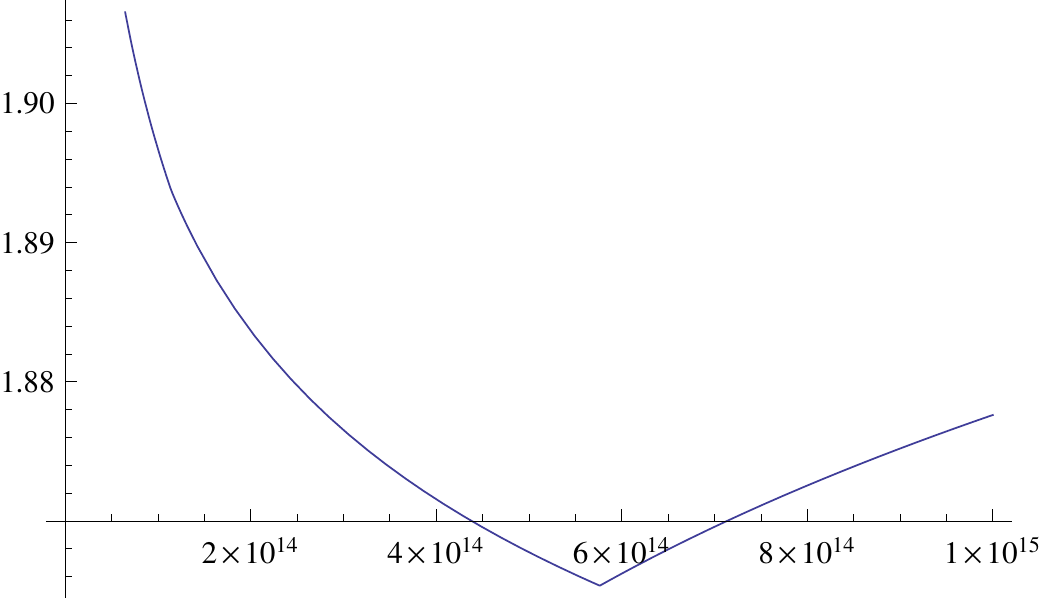}
\includegraphics[scale=0.5]{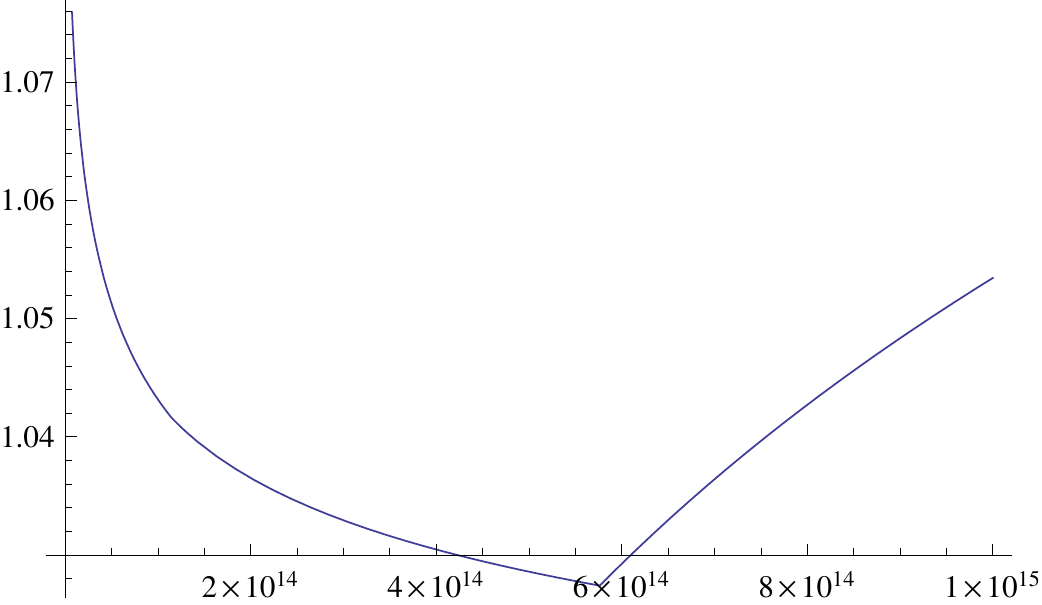}
\caption{Coefficients $\fa$ and $\fb$ as functions of the energy scale $\Lambda$ 
near the top see-saw scale.\label{abFig}}
\end{figure}

We concentrate here instead on the coefficients $\fa$, $\fb$, $\fc$, $\fd$, and $\fe$
of \eqref{abcde} and on their dependence on the energy scale $\Lambda$ through
their dependence on the Yukawa parameters and the Majorana mass terms and
the renormalization group equations \eqref{RGEYu}--\eqref{RGElambda}. The
renormalization group flow runs between the electroweak scale $\Lambda_{ew} = 
10^2$ GeV and the unification scale, chosen as in \cite{AKLRS} at $2\times 10^{16}$ GeV.

\begin{figure}
\includegraphics[scale=0.5]{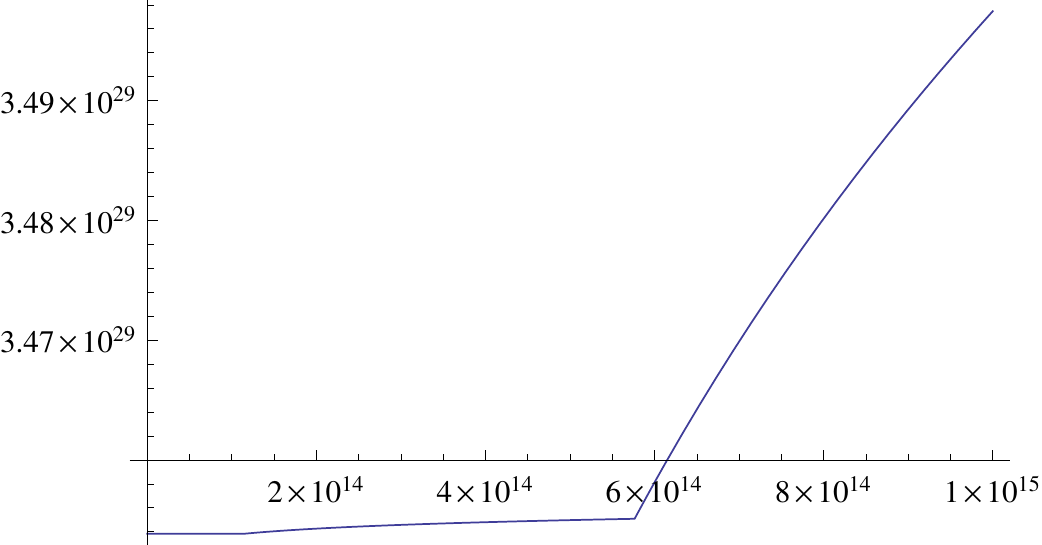}
\includegraphics[scale=0.5]{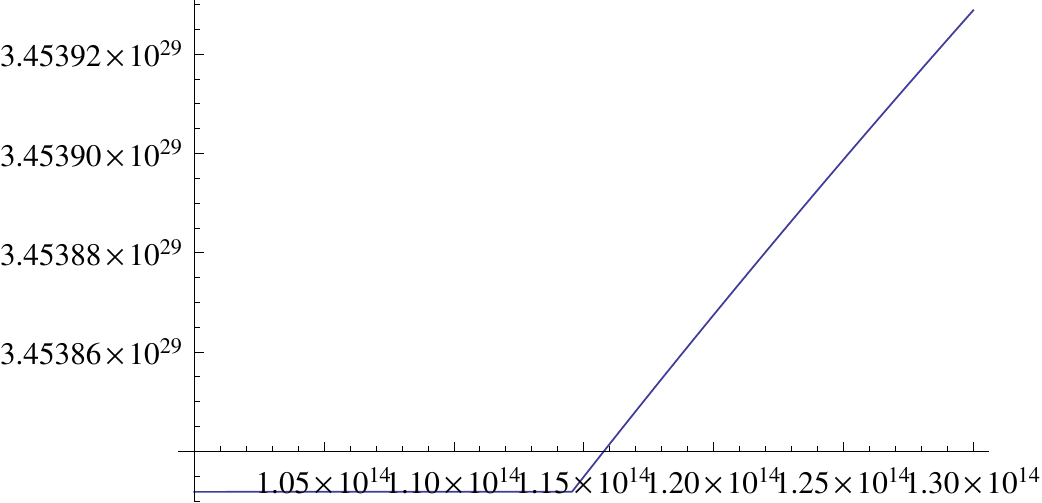}
\includegraphics[scale=0.5]{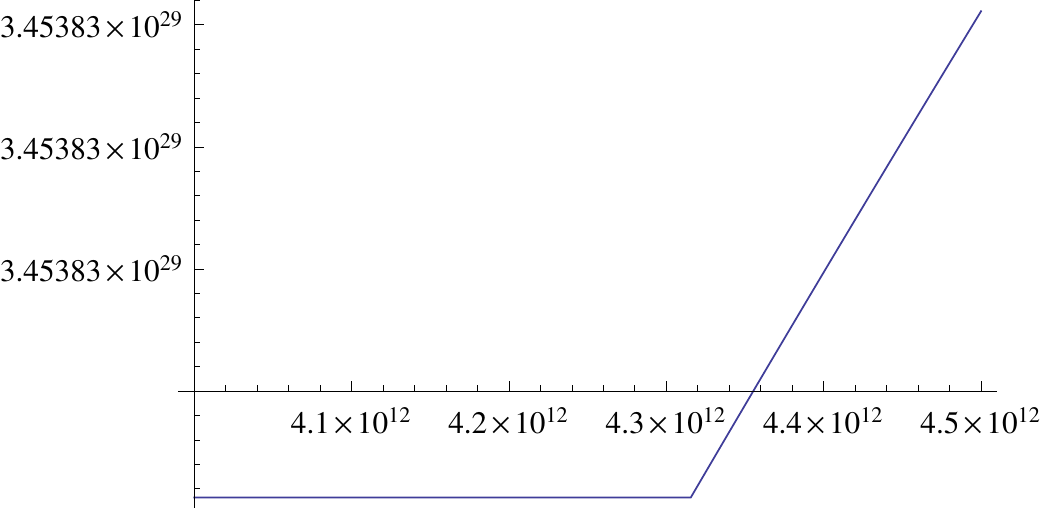}
\caption{The coefficient $\fc$ as a function of the energy scale $\Lambda$ near the three see-saw scales.\label{cFig}}
\end{figure}

By solving numerically the equations and plotting the running of the
coefficients \eqref{abcde}  one finds that the coefficients $\fa$ and $\fb$
show clearly the effect of the first (highest) see-saw scale, while the
effect of the two lower see-saw scales is suppressed. 

\begin{figure}
\includegraphics[scale=0.5]{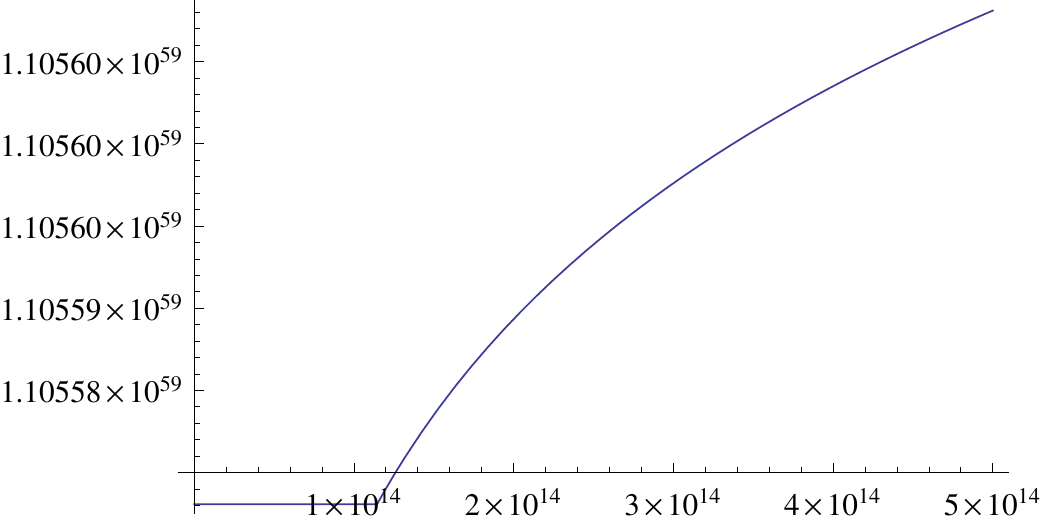}
\includegraphics[scale=0.5]{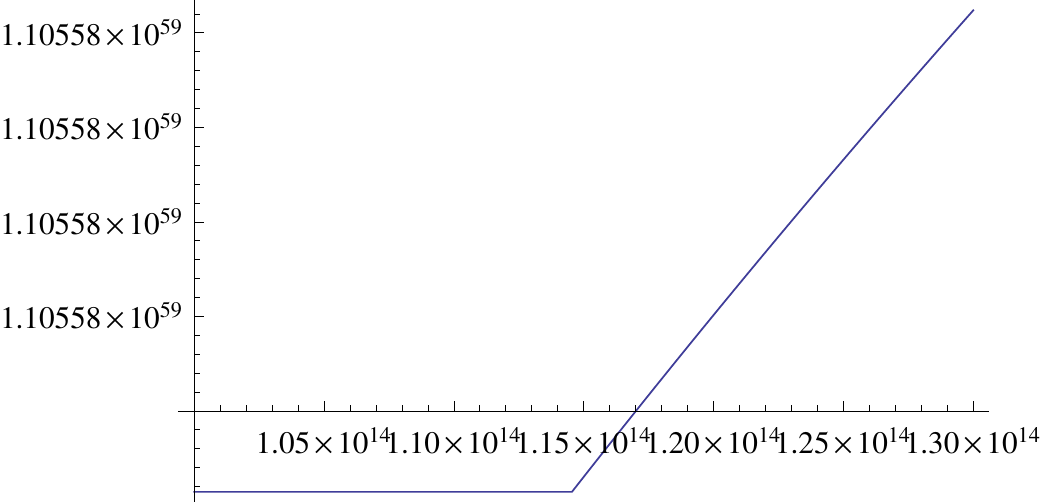}
\includegraphics[scale=0.5]{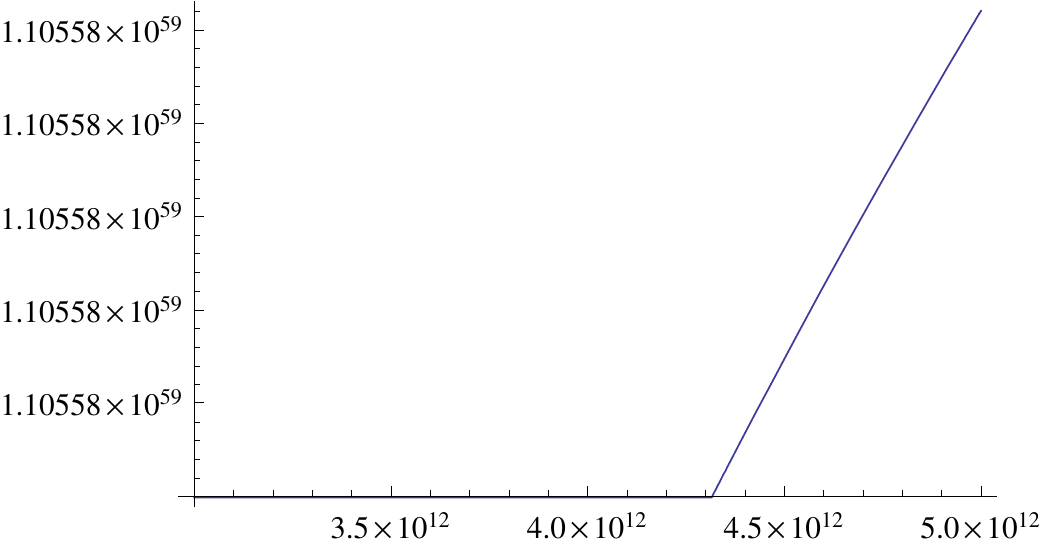}
\caption{The coefficient $\fd$ as a function of the energy scale $\Lambda$ near the three see-saw scales.\label{dFig}}
\end{figure}

The running of the coefficients $\fc$ and $\fd$ exhibits the effect of all three
see-saw scales, while the running of the remaining coefficient $\fe$ is the only
one that exhibits a large jump at the highest see-saw scale.

\begin{figure}
\includegraphics[scale=0.5]{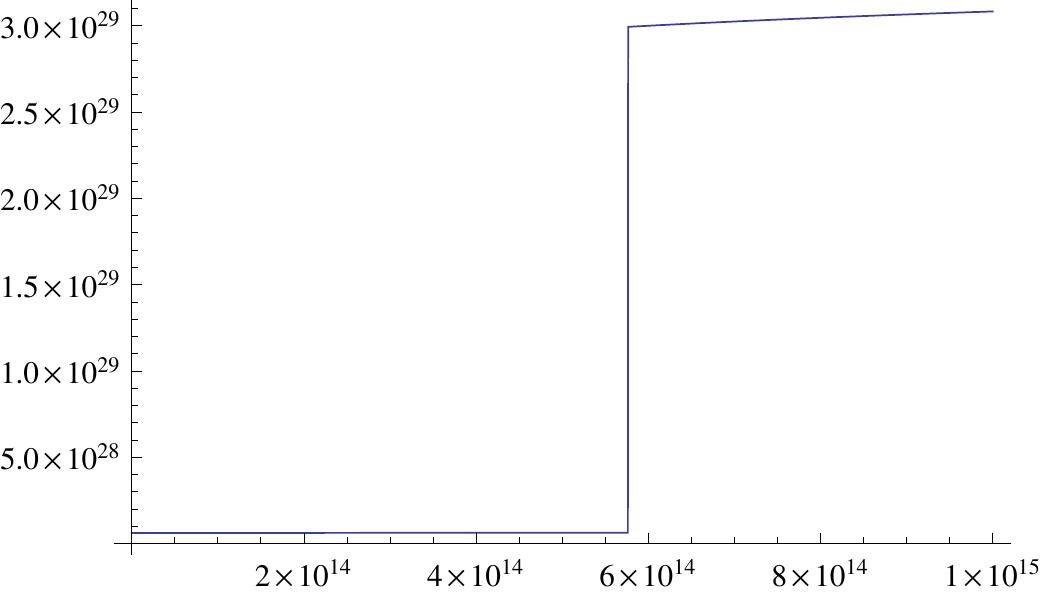}
\includegraphics[scale=0.5]{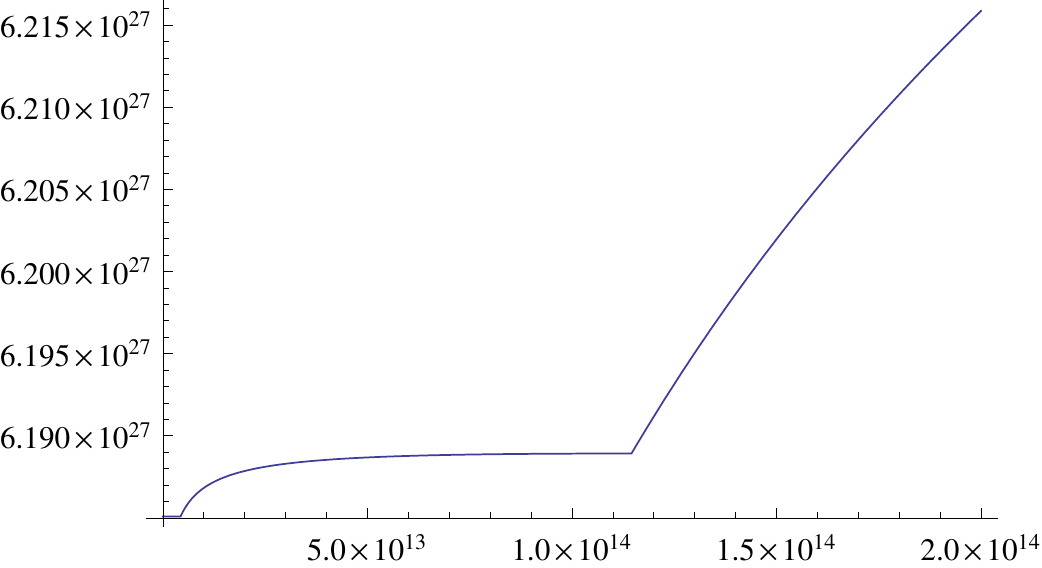}
\includegraphics[scale=0.5]{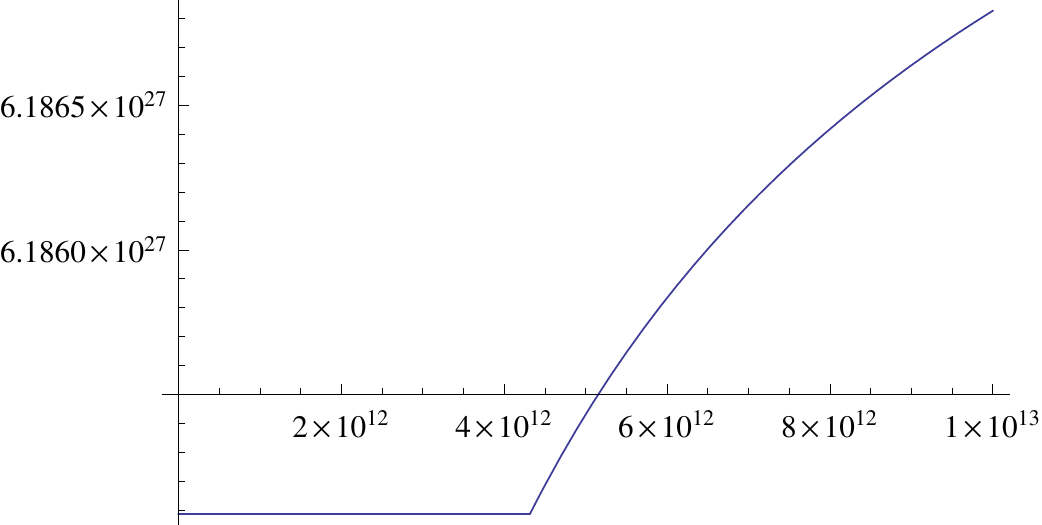}
\caption{The coefficient $\fe$ as a function of the energy scale $\Lambda$ near the three see-saw scales.\label{eFig}}
\end{figure}

Notice that the lack of differentiability at all the see-saw scales is inevitable, due to the
procedure used in \cite{AKLRS} and recalled here above for the construction of the 
effective field theories in between the different see-saw scales. 

\section{Cosmological implications of the model}\label{CosmoSec}

We now use the information on the running of the coefficients \eqref{abcde}
with the renormalization group to study the effect on the coefficients of the
gravitational and Higgs terms in the asymptotic expansion of the spectral
action. We derive some information about cosmological models of the
early universe that arise naturally in this noncommutative geometry setting.
In particular, we focus on the following aspects.

\begin{itemize}
\item Spontaneously arising Hoyle--Narlikar cosmologies in Einstein--Hilbert backgrounds.
\item Linde's hypothesis of antigravity in the early universe, via running of the
gravitational constant and conformal coupling to the Higgs.
\item Gravity balls from conformal coupling to the Higgs field. 
\item Detectable effects on gravitational waves of the running gravitational constant, as in
modified gravity theories.
\item Primordial black holes with or without gravitational memory.
\item Higgs based slow-roll inflation.
\item Varying effective cosmological constant and vacuum-decay.
\item Cold dark matter from Majorana masses of right handed neutrinos.
\end{itemize}

The main features of the noncommutative geometry model that will be discussed
in the following and that lead to the effects listed above are summarized as
follows.

\begin{itemize}
\item Variable effective gravitational constant.
\item Variable effective cosmological constant.
\item Conformal gravity.
\item Conformal coupling of the Higgs field to gravity.
\end{itemize}

\medskip

\subsection{Einsten gravity and conformal gravity}

The usual Einstein--Hilbert action (with cosmological term)
$$ \frac{1}{16 \pi G} \int R \sqrt{g} d^4 x + \gamma_0 \int \sqrt{g} d^4 x $$
minimally coupled to matter gives the Einstein field equations
$$ R^{\mu\nu} - \frac{1}{2} g^{\mu\nu} R + \gamma_0 g^{\mu\nu} = -8\pi G T^{\mu\nu}, $$
where the energy momentum tensor $T^{\mu\nu}$ is obtained from the matter part of
the Lagrangian, see \S 9.7 of \cite{CoMa} and \S  7.1.13 of \cite{Weinberg}. (Here we
use the notation $\gamma_0$ for the (variable) cosmological constant, which is more
frequently denoted by $\Lambda$ or $\lambda$ in the cosmology and general relativity
literature, but unfortunately both of these letters are already assigned other meanings here.)
In addition to these terms, where both $G$ and $\gamma_0$ will be running with the
energy scale $\Lambda$ and depending on the free parameters $f_2$ and $f_4$,
the asymptotic expansion for the spectral action also delivers conformal gravity terms.
Conformal gravity is considered an alternative to the usual form of general relativity,
where the Einstein--Hilbert action is replaced by an action based on the Weyl curvature tensor
$$ C_{\lambda\mu\nu\kappa} = R_{\lambda \mu\nu\kappa} -\frac{1}{2} (g_{\lambda\nu} R_{\mu\kappa}
-g_{\mu\nu} R_{\lambda\kappa} + g_{\mu\kappa} R_{\lambda\nu}) + \frac{1}{6} (g_{\lambda\nu}
g_{\mu\kappa} - g_{\lambda\kappa}g_{\mu\nu}). $$
This has the property of being conformally invariant, namely under a transformation of the
form $g_{\mu\nu}(x) \mapsto f(x)^2 \, g_{\mu\nu}(x)$ the Weyl tensor remains unchanged,
$C_{\lambda\mu\nu\kappa}  \mapsto C_{\lambda\mu\nu\kappa}$. The action functional of
conformal gravity is of the form
\begin{equation}\label{confact}
 \alpha_0 \int C_{\lambda\mu\nu\kappa} C^{\lambda\mu\nu\kappa} \sqrt{g} d^4x , 
\end{equation} 
upon rewriting the above in terms of the Riemann curvature tensor
$$ \alpha_0 \int (R_{\lambda\mu\nu\kappa} R^{\lambda\mu\nu\kappa} - 2 R_{\mu\nu} R^{\mu\nu} +
\frac{1}{3} R^2) \sqrt{g} d^4x $$
and using the fact that $R_{\lambda\mu\nu\kappa} R^{\lambda\mu\nu\kappa} - 4 R_{\mu\nu} R^{\mu\nu} 
+R^2$ is a total divergence (see \cite{Mann}) one can rewrite the conformal action functional as 
$$ 2\alpha_0 \int (R_{\mu\nu} R^{\mu\nu} -\frac{1}{3} R^2) \sqrt{g} d^4x $$
which gives field equations
$$ W^{\mu\nu} = -\frac{1}{4\alpha_0} T^{\mu\nu}, $$
where
$$ W^{\mu\nu} =   2 C^{\mu\lambda\nu\kappa}_{;\nu;\kappa} - C^{\mu\lambda\nu\kappa} R_{\lambda\kappa} = $$ $$ \frac{1}{6} g_{\mu\nu} \nabla^2 R + \frac{1}{3} \nabla_\mu \nabla_\nu R -
\frac{1}{3} R (2 R_{\mu\nu} - \frac{1}{2} g_{\mu\nu} R) + R^{\beta\rho} (\frac{1}{2} g_{\mu\nu} R_{\beta\rho} - 2 R_{\beta\mu\rho\nu}) . $$

In addition to the Weyl curvature tensor itself being conformally invariant, one can 
add to the conformal gravity action a coupling to a field $\varphi$. Under a conformal
transformation $g_{\mu\nu}(x) \mapsto f(x)^2 \, g_{\mu\nu}(x)$, a field transforming like
$\varphi \mapsto f^{-1} \varphi$ gives
$$ (\partial_\mu \varphi)^2 \mapsto f^{-4}((\partial_\mu \varphi)^2 +\varphi (\varphi \nabla_\mu
\log f - 2 \nabla_\mu \varphi) \nabla^\mu \log f), $$
while the scalar curvature transforms like
$$ R \mapsto f^{-2} (R - (d-1)((d-2) \nabla_\mu \log f \nabla^\mu \log f + 2 \nabla^\mu\nabla_\mu \log f) $$
so a non-minimal coupling of the field $\varphi$ to gravity of the form
$$ 2\xi_0 \int R \varphi^2 \sqrt{g} d^4 x $$
is conformally invariant in dimension $d=4$ if $2\xi_0 = 1/6$. In addition to these terms a
quartic potential 
$$ \lambda_0 \int \varphi^4 \sqrt{g} d^4 x $$
also preserves conformal invariance.
Thus, adding to the conformal action \eqref{confact} terms of the form
\begin{equation}\label{confphi}
 \frac{1}{2} \int (\partial_\mu \varphi)^2 \sqrt{g} d^4 x - \frac{1}{12} \int R \varphi^2 \sqrt{g} d^4x +\lambda_0 \int \varphi^4 \sqrt{g} d^4 x 
\end{equation} 
maintains the conformal invariance. 
The conformal gravity action \eqref{confact} with an additional non-minimal conformal
coupling to another field $\varphi$ in the form \eqref{confphi} is the basis of the
Hoyle--Narlikar cosmologies, which were proposed as possible models for steady state
cosmologies in \cite{HoNa}, with the field $\varphi$ related to the Mach principle. 
In fact, the presence of the field $\varphi$ allows for a modification of the energy
momentum tensor of the form $T^{\mu\nu} \mapsto T^{\mu\nu} - \xi_0 \varphi^2 (R^{\mu\nu} -\frac{1}{2} g^{\mu\nu} R) - g^{\mu\nu} \lambda_0 \varphi^4$, which was used as ``creation field" in steady state models. While the steady state cosmologies fail to account for major cosmological phenomena
such as the background radiation, hence Hoyle--Narlikar cosmologies cannot be extrapolated 
towards the early universe, conformal gravity remains a valuable model (see \cite{Mann} for
a recent discussion). Within the noncommutative geometry model we will see below that
one typically has a dominant Einstein--Hilbert action, and only at certain scales where the
behavior of the running effective gravitational constant presents phase transitions one finds
that the subdominant terms of conformal gravity become dominant. This gives rise to
emergent Hoyle--Narlikar cosmologies, for which the problem of extrapolating towards 
earlier times does not arise, as they become suppressed by the dominant Einstein--Hilbert 
term away from the energy scale of the phase transition. In the noncommutative geometry
model the role of the field non-minimally conformally coupled to gravity is played by the Higgs field.
Variants of the model of \cite{CCM}, discussed for instance in \cite{Cham}, allow for the presence of 
a further scalar field $\sigma$ also conformally non-minimally coupled to $R$. Most of those
arguments we describe in the following sections that are based on the coupling of the Higgs to gravity
can be formulated also in terms of this other field $\sigma$, though we will not explicitly mention it.

\subsection{Variable effective gravitational constant}

According to the expressions for the coefficients \eqref{coeffsrun}
of the terms in the asymptotic expansion
of the spectral action, we see that this model has an Einstein--Hilbert term, where the usual
coefficient $\frac{1}{16 \pi G} \int R \, \sqrt{g} d^4 x$, with $G$ the Newton constant $G\sim (10^{19} {\rm GeV})^{-2}$, with $1/\sqrt{G}=1.22086 \times 10^{19}$ GeV 
the Planck mass, is replaced by an effective gravitational constant of the form
\begin{equation}\label{effNewton}
 G_{{\rm eff}}= \frac{\kappa_0^2}{8\pi} = \frac{3\pi}{192 f_2 \Lambda^2 - 2 f_0 \fc(\Lambda)} 
\end{equation} 
In this expresstion, the parameter $f_0$ is fixed by the unification condition \eqref{f03ways}
(we will use the first value for simplicity), the parameter $f_2$ is unconstrained in $\R^*_+$
and the function $\fc(\Lambda)$ is determined by the renormalization group equations.

Thus, one can best represent the effective gravitational constant of the model as
a surface $G_{\rm eff}(\Lambda,f_2)$, which is a function of the energy scale 
in the range $\Lambda_{ew} \leq \Lambda \leq \Lambda_{unif}$ between the
electroweak and the unification scales. It is in fact often preferable to consider the
surface $G_{\rm eff}^{-1}(\Lambda,f_2)$ that describes the inverse effective 
gravitational constant, since it is in this form that it appears in the coefficients of
the action functional, and it does not have singularities.

\begin{figure}
\includegraphics[scale=0.8]{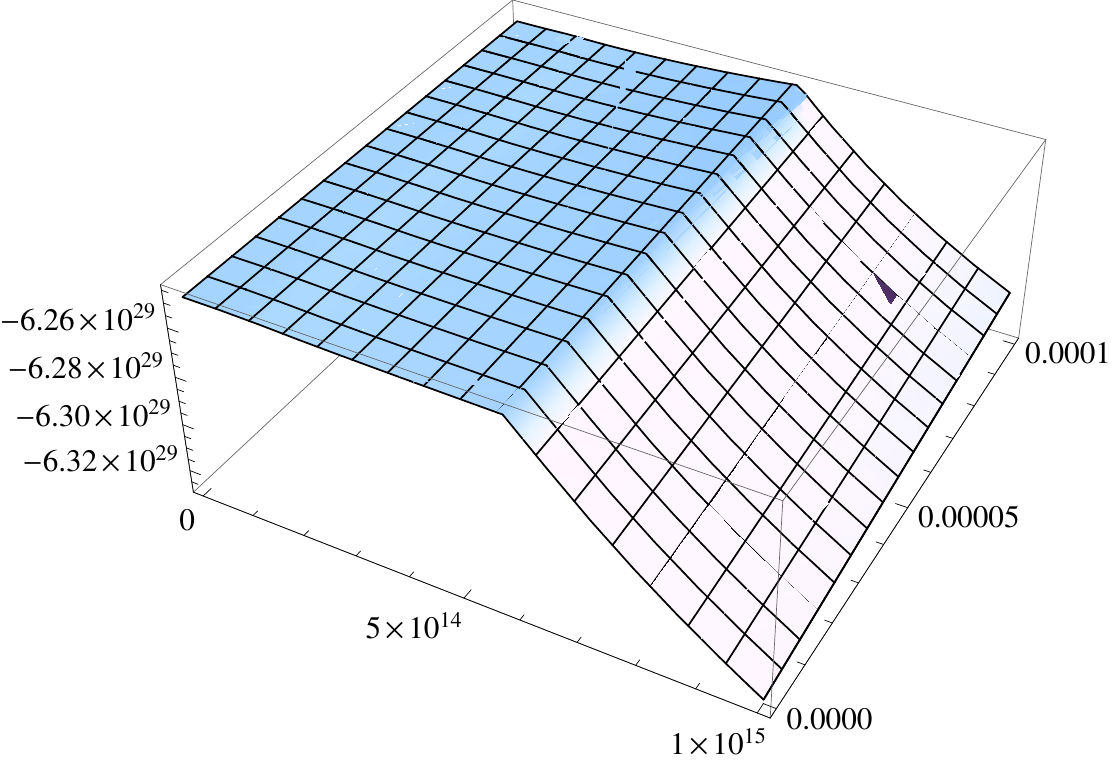}
\caption{The region of the surface $G_{\rm eff}(\Lambda,f_0)^{-1}$ in the range
$10^{14} \leq \Lambda \leq 10^{15}$ GeV and for $10^{-16}\leq f_2\leq 10^{-4}$.
\label{GeffinvFig}}
\end{figure}

As an example of the form of the surface $G_{\rm eff}^{-1}(\Lambda,f_2)$, one
sees in Figure \ref{GeffinvFig} 
that, for sufficiently small $f_2$, the surface exhibits a kink at the top see-saw scale.

As another example, if one requires in this model that the value of the
effective gravitational constant at the lower end of the energy spectrum we
are considering, that is, at the electroweak scale, already agrees with the
usual Newton constant, this requires a very large fine tuning of the parameter 
$f_2$, which is fixed to have the value $ f_2= 7.31647\times 10^{32}$. 
Notice that this is just an example. There is no physical reason to assume in
this model that at the time of the electroweak epoch of the early universe the
effective gravitational constant would have to be already equal to that of the
modern universe. In this example one then sees that the running of 
$G_{\rm eff}^{-1}$ becomes dominated entirely by the term $192 f_2 \Lambda^2/(3\pi)$,
with the term $2 f_0 \fc(\Lambda)/(3\pi)$ remaining all the while several orders of magnitude smaller.
This is in contrast with the running of Figure \ref{GeffinvFig} where $f_2$ is small,
and the running of $2 f_0 \fc(\Lambda)/(3\pi)$ comes to play a significant role, as one sees from
the effect of the see-saw scales.

\subsection{Emergent Hoyle--Narlikar cosmologies}

We look more closely at cases with large $f_2$. For simplicity we illustrate
what happens in this range by focusing on the
example we mentioned above where $f_2$ is fixed so that 
$G_{\rm eff}(\Lambda_{ew})=G$, though qualitatively the results described in this
section hold for a wider range of choices of $f_2$ sufficiently large.
We show that, in the case of spaces with $R\sim 1$,
a phase transition happens at the top see-saw scale, where the dominant 
Einstein--Hilbert action is suppressed and the action of the system is 
dominated by a spontaneously arising Hoyle--Narlikar type cosmology.

We first identify the dominant terms in the action, under the hypothesis
of large $f_2$. Then we identify the conditions under which these 
terms are suppressed and the remaining terms become dominant.

\begin{prop}\label{dominantprop}
For sufficiently large values of the parameter $f_2$ (for instance when 
$G_{\rm eff}(\Lambda_{ew})=G$), and in the range of energies where
the effective cosmological term is kept small by the choice of the
parameter $f_4$, the dominant terms in the 
expansion of the spectral action \eqref{SAvarchange} are 
\begin{equation}\label{dominant1st}
\Lambda^2 \left( \frac{1}{2\tilde\kappa_0^2} \int R \sqrt{g} d^4x - 
\tilde\mu_0^2 \int |H|^2 \sqrt{g} d^4x \right) ,
\end{equation}
for $\tilde\kappa_0 = \Lambda \kappa_0$ and $\tilde\mu_0 =\mu_0/\Lambda$. 
\end{prop}

\proof
In this case, as we have seen already, the running of the effective gravitational
constant is dominated by the term $G_{\rm eff}^{-1}\sim 192 f_2 \Lambda^2/(3\pi)$.
Similarly, in the quadratic term of the Higgs $\mu_0^2$, the term  $2 f_2 \Lambda^2/f_0$ 
is dominant when $f_2$ is sufficiently large. In particular, in the example where
$G_{\rm eff}(\Lambda_{ew})=G$, the term $2 f_2 \Lambda^2/f_0$ satisfies
$$ \frac{2 f_2 \Lambda_{ew} ^2}{f_0} = 1.71627\times 10^{36}, $$
while the second term is several orders of magnitude smaller,
$$ -\frac{\fe(\Lambda_{ew} )}{\fa(\Lambda_{ew} )} = -1.51201 \times 10^{27}, $$
even though the coefficient $\fe$ varies more significantly than the other
coefficients of \eqref{abcde}.

We then proceed to estimate the remaining terms of \eqref{SAvarchange} and
show that they are all suppressed with respect to the dominant terms above, for
this choice of $f_2$.
The parameters $\alpha_0$, $\tau_0$, $\xi_0$ are not running with $\Lambda$,
and we can estimate them to be 
$$ \alpha_0 \sim - 0.25916, \ \ \ \tau_0 \sim 0.158376, $$
while $\xi_0=1/12$ remains fixed at the conformal coupling value.

\begin{figure}
\includegraphics[scale=0.5]{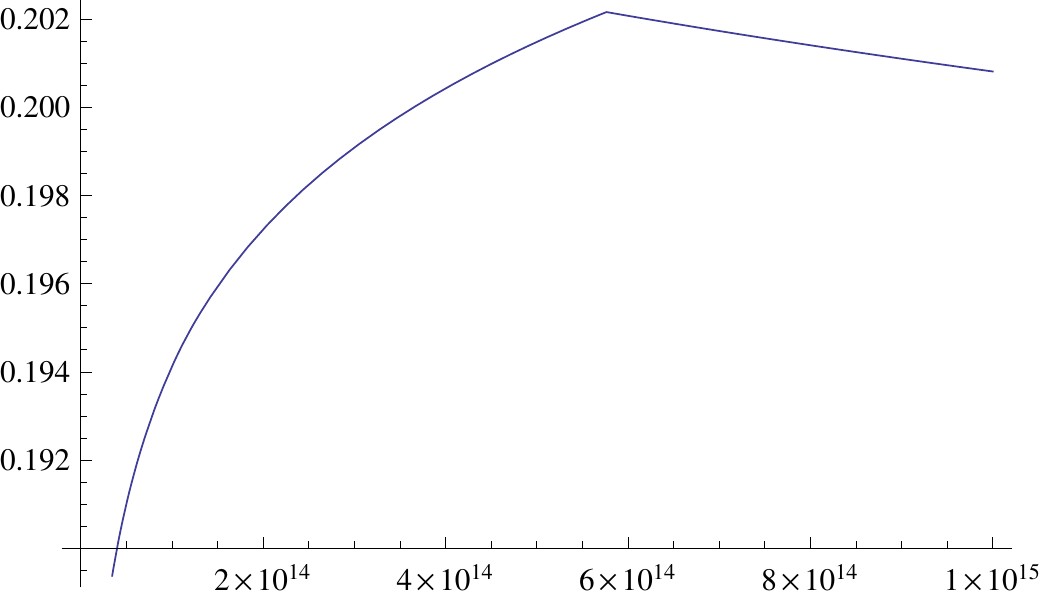}
\includegraphics[scale=0.5]{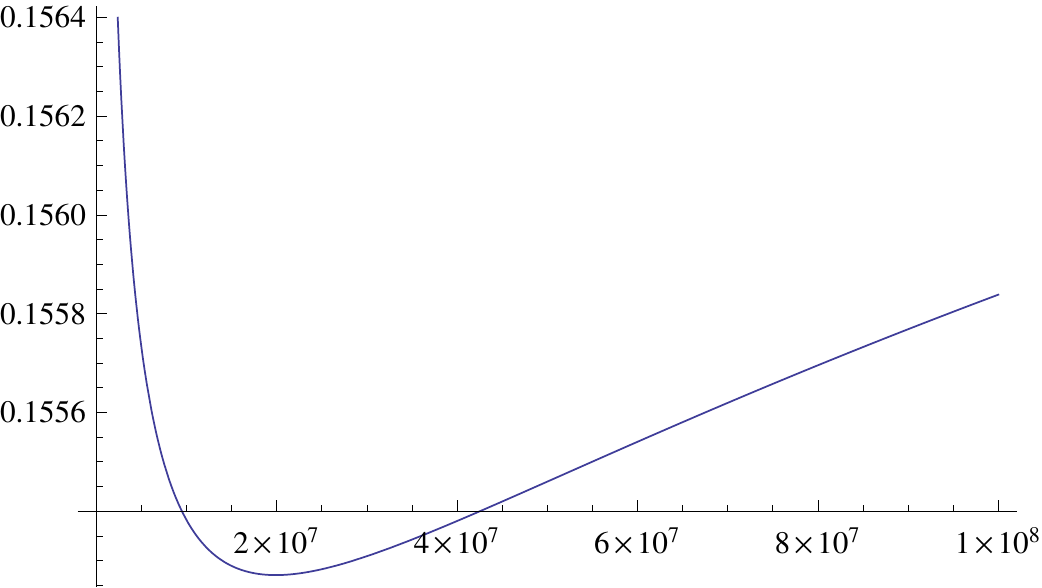}
\caption{The running of $\lambda_0$ near the top see-saw scale and near the
electroweak scale.\label{lambda0runFig}}
\end{figure}

To estimate the running of the coefficient $\lambda_0$, we propose an
{\em ansatz} on how it is related to the running of $\lambda$ in the RGE \eqref{RGElambda}
and to that of the coefficients $\fa$ and $\fb$.  
We also know that the boundary conditions at unification of \eqref{coeffsrun} satisfy
$$ \lambda_0 |_{\Lambda=\Lambda_{unif}}=  \lambda(\Lambda_{unif}) \frac{\pi^2 \fb(\Lambda_{unif}) }{ f_0 \fa^2(\Lambda_{unif}) } , $$
where in \eqref{RGElambda} one uses the boundary condition $\lambda(\Lambda_{unif})=1/2$
as in \cite{AKLRS}. Thus, we investigate here the possibility that the coefficient $\lambda_0$
runs with the RGE according to the relation
\begin{equation}\label{lambdaRun}
 \lambda_0(\Lambda)= \lambda(\Lambda) \frac{\pi^2 \fb(\Lambda) }{ f_0 \fa^2(\Lambda) } .
\end{equation}
With this {\em ansatz} for the running of $\lambda_0$ one finds that the value of
$\lambda_0$ varies between $\lambda_0(\Lambda_{ew})=0.229493$ and
$\lambda_0(\Lambda_{unif})= 0.184711$, along the curve shown in Figure \ref{lambda0runFig},
with a maximum at $\lambda_0(\Lambda_{ew})$ and a local maximum at the top see-saw scale
with value $0.202167$, and with a minimum near $2\times 10^7$ GeV of value $0.155271$. We discuss this {\em ansatz} more in details in \S \ref{ansatzSec} below. 

Finally, for a given energy scale $\Lambda$ it is always possible to eliminate the cosmological
term at that energy scale by adjusting the coefficient $f_4$  (see \S \ref{cosmoconstSec}). 
The vanishing condition for the effective cosmological constant at energy $\Lambda$ is 
realized by the choice of
\begin{equation}\label{f4Lambda}
 f_4 = \frac{(4 f_2 \Lambda^2 \fc - f_0 \fd )}{192 \Lambda^4} .
\end{equation}

Thus, if at a given energy scale the coefficient $f_4$ is chosen so that the effective
cosmological constant vanishes, then the only terms that remain as dominant terms
in the action are those of \eqref{dominant1st}.
\endproof

In this scenario then a new feature arises. Namely, the fact that one has comparable terms 
$$ \frac{1}{2\kappa_0^2}\sim \frac{96 f_2 \Lambda^2}{24 \pi^2} \sim  2.96525 \times 10^{32} 
\Lambda^2 $$
and
$$ \mu_0^2 \sim \frac{2 f_2 \Lambda^2}{f_0} \sim 1.71627 \times 10^{32} \Lambda^2 $$
can lead to cancellations under suitable geometric hypotheses.

\begin{prop}\label{HNcosmo}
Consider a space with $R\sim 1$. Then, for values $|H|\sim \sqrt{\fa f_0}/\pi$,
the term \eqref{dominant1st}, which is the dominant term for $f_2$ sufficiently large,
vanishes and is replaced by the sub-dominant
\begin{equation}\label{subdom}
\frac{\fe}{\fa} |H|^2 - \frac{f_0 \fc}{24 \pi^2} R \sim
\frac{f_0}{\pi^2}(\fe - \frac{\fc}{24})
\end{equation}
as the leading term in the formula \eqref{SAvarchange} for the spectral action.
Near the top see-saw scale, the term \eqref{subdom}
has a discontinuity, where the dynamics becomes dominated by a
Hoyle--Narlikar cosmology given by the remaining terms of  \eqref{SAvarchange}.
\end{prop}

\proof As in \cite{CoMa},
Corollary 1.219, we expand the Higgs field around $|H| \sim  \sqrt{\fa f_0}/\pi$. 
Then, one can compare the two terms
\begin{equation}\label{constRcase}
 \frac{1}{2\tilde\kappa_0^2} R  - \tilde\mu_0^2  |H|^2 .
\end{equation} 
This identifies a value for the constant curvature
$$ R \sim \frac{2\tilde\kappa_0^2 \tilde\mu_0^2 \fa f_0}{\pi^2} $$
at which the dominant term 
$$ \frac{96 f_2 \Lambda^2}{24 \pi^2} R - \frac{2 f_2 \Lambda^2}{f_0}  \frac{\fa f_0}{\pi^2} $$
of \eqref{constRcase} vanishes, leaving the smaller terms to dominate the dynamics.
One can estimate that this gives a value for the scalar curvature very close to one,
$R = 0.979907 \sim 1$, if we use the value of $\fa$ at unification energy and the first 
possible value of $f_0$ in \eqref{f03ways}.

The next smaller term in \eqref{constRcase}  is then of the form
$$ \frac{\fe}{\fa} |H|^2 - \frac{f_0 \fc}{24 \pi^2} R \sim \frac{\fe}{\fa} \frac{\fa f_0}{\pi^2} - \frac{f_0 \fc}{24 \pi^2}  $$
which gives \eqref{subdom}. 
Near the top see-saw scale, at around $5.76405 \times 10^{14}$ GeV the term
$\fe-\fc/24$ has a jump and a sign change due to the large jump of the coefficient
$\fe$ near the top see-saw scale (see Figure \ref{eFig}).
At this phase transition what is left of the dynamics of \eqref{SAvarchange} 
are the remaining terms. One therefore sees an emergent behavior where
near the phase transition of the top see-saw scale, of the following form.

The coefficient $f_4$ can be chosen so that the cosmological term
vanishes at this same top see-saw scale energy (see \S \ref{cosmoconstSec}). 
The dynamics of the
model is then dominated by the remaining terms, which recover a well known treatment 
of gauge and Higgs field in {\em conformal gravity}, as discussed for instance in \S 2.2 of
\cite{EFP}. According to these models, the conformally invariant action for the gauge
and Higgs bosons is given by the terms
\begin{equation}\label{gaugeHconf}
\begin{array}{ll}
S_c =& \displaystyle{ \alpha_0 \int C_{\mu
\nu \rho \sigma} \, C^{\mu \nu \rho \sigma} \sqrt g \,d^4 x}
+  \displaystyle{ \frac{1}{2} \int\,  |D H|^2\, \sqrt g \,d^4 x }  \\[3mm] 
&- \displaystyle{ \xi_0 \int R \, |H|^2 \, \sqrt g \,d^4 x} 
+  \displaystyle{\lambda_0 \int |H|^4 \, \sqrt g \,d^4 x } \\[3mm]
&+ \displaystyle{ \frac{1}{4} \int\,(G_{\mu \nu}^i \, 
G^{\mu \nu i} +  F_{\mu
\nu}^{ \alpha} \, F^{\mu \nu  \alpha}+\, B_{\mu \nu} \, B^{\mu \nu})\, \sqrt g \,d^4 x }.
\end{array}
\end{equation}
The additional topological term $\tau_0 \int R^* R^* \sqrt g \,d^4 x$ is non-dynamical
and only contributes the Euler characteristic of the manifold. The action \eqref{gaugeHconf}
is that of a Hoyle--Narlikar cosmology.
\endproof

In this
model, the scalar curvature $R$, which is constant or near constant, provides an effective
quadratic term for the Higgs and a corresponding symmetry breaking phenomenon, as
observed in \cite{EFP}. This produces in turn a breaking of conformal symmetry, via the
Higgs mechanism giving mass to some of the gauge field, thus breaking conformal
invariance. In this range, with a constant curvature $R$ and in the absence of the
quadratic term in $\mu_0$, the Higgs field is governed by a potential of the form
$\cV_{R=1}(H)= -\xi_0 |H|^2 + \lambda_0 |H|^4$, which has a minimum at
$|H|^2=\xi_0/(2\lambda_0)$. 

We discuss in \S \ref{ConfGravSec} below another instance of emergent
Hoyle--Narlikar cosmology based on conformal gravity at a phase transition
where the effective gravitational constant undergoes a sign change and an
antigravity regime arises.

\subsection{Effects on gravitational waves}\label{gravwaveSec}

The fact that we have a variable gravitational constant in the model has detectable
effects on phenomena like gravitational waves whose propagation depends on the
value of the gravitational constant. 

Under the assumptions that the remaining terms in
the asymptotic expansion of the spectral action are negligible with respect to the 
dominant \eqref{dominant1st}, and further that  $|H|\sim 0$, so that only
the Einstein--Hilbert term dominates, one can show as in \cite{NeSa} that
the equations of motion for \eqref{SAvarchange} reduce to just
$$ R^{\mu\nu}- \frac{1}{2} g^{\mu\nu} R = \kappa_0^2 T^{\mu\nu}. $$
In \cite{NeSa} the authors conclude from this that, in the isotropic case, 
the noncommutative geometry model has no effect on the gravitational waves
that distinguish it from the usual Einstein--Hilbert cosmology.
However, in fact, even in this case one can find detectable effects on the
gravitational waves that distinguish the noncommutative geometry model from the 
ordinary case of general relativity, because of the running of the effective 
gravitational constant.

We consider here two different scenario for the time-energy relation, one
which will be relevant close to the electroweak  scale (\cite{Guth}), where 
$\Lambda \sim t^{-1/2}$, and the other that refers to the inflationary period,
with $\Lambda \sim e^{-\alpha t}$. We show that in both cases the behavior
of the gravitational waves differs from the behavior, with the same time-energy
conversion, of the solutions in the classical case, thus detecting the presence
of noncommutative geometry.

\begin{prop}\label{gravwaveprop}
In between the unification and the electroweak scale the gravitational waves
propagate according to the $\Lambda$-dependent equation
\begin{equation}\label{friedeqLambda}
  - 3 \left(\frac{\dot{a}}{a}\right)^2 + \frac{1}{2} \left( 4 \left(\frac{\dot{a}}{a}\right) \dot{h} + 2 \ddot{h} \right)
  =  \frac{12\pi^2}{96 f_2 \Lambda^2 -  f_0 \fc(\Lambda)} \, T_{00}.
\end{equation}  
Upon rewriting the energy variable $\Lambda$ as a function of time through $\Lambda=1/a(t)$, 
one obtains
\begin{equation}\label{wavesLambda1}
\ddot{h} + t^{-1} \dot{h} -\frac{3}{4} t^{-2} = \frac{12\pi^2 T_{00}}{96 f_2 t^{-1} -  f_0 \fc(t^{-1/2})}.
\end{equation} 
in the radiation dominated era where $\Lambda \sim t^{-1/2}$, and 
\begin{equation}\label{wavesLambda2}
\ddot{h} +2 \alpha \dot{h} -3 \alpha^2 = \frac{12\pi^2 T_{00}}{96 f_2 e^{-2\alpha t} -  f_0 \fc(e^{-\alpha t})}.
\end{equation} 
in the inflationary epoch where $\Lambda \sim e^{-\alpha t}$.
\end{prop}

\proof For a metric of the form
\begin{equation}\label{atmetric}
 g_{\mu\nu} = a(t)^2 \left(\begin{array}{cc} -1 & 0 \\ 0 & \delta_{ij}+ h_{ij}(x) \end{array}\right) 
\end{equation} 
one separates the perturbation $h_{ij}$ into a trace and traceless part, and 
the gravitational waves are then governed by the Friedmann equation, which gives
\begin{equation}\label{friedeq}
  - 3 \left(\frac{\dot{a}}{a}\right)^2 + \frac{1}{2} \left( 4 \left(\frac{\dot{a}}{a}\right) \dot{h} + 2 \ddot{h} \right)
  = \kappa_0^2 \, T_{00}.
\end{equation}  
This equation is {\em formally}
the same as the usual equation for the gravitational waves, up to replacing $\kappa_0^2$
for $8\pi G$, as remarked in \cite{NeSa}. However, the dependence of $\kappa_0^2$ on
the energy scale $\Lambda$ leads to the result \eqref{friedeqLambda}. 

The change of variable between energy and time, for a cosmology of the form \eqref{atmetric}
is given by $\Lambda = 1/a(t)$. Thus, we can write the equation \eqref{friedeq} by expressing the
right hand side also as a function of time in the form
\begin{equation}\label{wavesLambda}
- 3 \left(\frac{\dot{a}(t)}{a(t)}\right)^2 + \frac{1}{2} \left( 4 \left(\frac{\dot{a}(t)}{a(t)}\right) \dot{h}(t) + 2 \ddot{h}(t) \right) = \frac{12\pi^2 T_{00}}{96 f_2 \frac{1}{(a(t))^2} -  f_0 \fc(\frac{1}{a(t)})}.
\end{equation} 

In the radiation dominated era, where the function $a(t)$ behaves like $a(t)=t^{1/2}$ we find
$$ \frac{\dot{a}}{a} = \frac{1}{2} t^{-1}, $$
which gives
$$ - 3 \left(\frac{\dot{a}}{a}\right)^2 + \frac{1}{2} \left( 4 \left(\frac{\dot{a}}{a}\right) \dot{h} + 2 \ddot{h} \right) = \ddot{h} + t^{-1} \dot{h} -\frac{3}{4} t^{-2} . $$
This gives \eqref{wavesLambda1}. In the inflationary era, where the function $a(t)$ behaves 
exponentially $a(t)=e^{\alpha t}$, one obtains instead $\dot{a}/a=\alpha$ and
$$  - 3 \left(\frac{\dot{a}}{a}\right)^2 + \frac{1}{2} \left( 4 \left(\frac{\dot{a}}{a}\right) \dot{h} + 2 \ddot{h} \right)
= \ddot{h} +2 \alpha h -3 \alpha^2, $$
which gives \eqref{wavesLambda2}.
\endproof

For a choice of the parameter $f_2$ sufficiently large (such as the one that gives $G_{\rm eff}(\Lambda_{ew})=G$) for  which $\kappa_0^2(\Lambda)\sim \tilde\kappa_0^2/\Lambda^2$ 
we then obtain the following explicit solutions to the equation \eqref{friedeqLambda}.

\begin{prop}\label{gravWaveshort}
We consider the case with the parameter $f_2$ sufficiently large (for instance in the example
where $G_{\rm eff}(\Lambda_{ew})=G$).
In the radiation dominated era, where the metric \eqref{atmetric} has $a(t)\sim t^{1/2}$, 
and the energy-time relation is given by $\Lambda =1/a(t)$ the equation \eqref{wavesLambda1}
has solutions of the form
\begin{equation}\label{wavesLamSol1}
h(t) = \frac{4 \pi^2 T_{00}}{288 f_2} t^3 + B + A \log(t) + \frac{3}{8} \log(t)^2
\end{equation}
In the inflationary epoch, for a metric of the form \eqref{atmetric} with $a(t)\sim e^{\alpha t}$, 
for some $\alpha >0$,  the equation \eqref{wavesLambda2} has solutions of the form
\begin{equation}\label{wavesLamSol2}
 h(x) =  \frac{3 \pi^2 T_{00}}{192 f_2 \alpha^2} e^{2\alpha t} + \frac{3\alpha}{2} t +
 \frac{A}{2\alpha} e^{-2\alpha t} + B.
\end{equation}
\end{prop}

\proof
We assume that the parameter $f_2$ is sufficiently large. For example, we take the case
where $f_2$ is fine tuned so as to have an agreement between the value at the electroweak scale of the
effective gravitational constant and the usual Newton constant, $\kappa_0^2(\Lambda_{ew} ) = 8\pi G$. 
One then finds at high energies a different effective Newton constant, which behaves like
$$ \kappa_0^2 \sim \frac{12 \pi^2}{96 f_2} \Lambda^{-2}, $$
which then gives a modified gravitational waves equation
\begin{equation}\label{friedeqLambda2}
  - 3 \left(\frac{\dot{a}}{a}\right)^2 + \frac{1}{2} \left( 4 \left(\frac{\dot{a}}{a}\right) \dot{h} + 2 \ddot{h} \right)
  =  \frac{\tilde\kappa_0^2}{\Lambda^2}\, T_{00},
\end{equation}  
with $\tilde \kappa_0^2 =\frac{12 \pi^2}{96 f_2}$.

We look first at the radiation dominated case where $a(t)=t^{1/2}$. In the interval of
energies that we are considering, this has relevance close to the electroweak scale, 
see \cite{Guth}. We then have equation
\eqref{wavesLambda1} in the form
\begin{equation}\label{wavesLambda1a}
\ddot{h} + t^{-1} \dot{h} -\frac{3}{4} t^{-2} = t \, \frac{12\pi^2 T_{00}}{96 f_2}.
\end{equation}
Assuming $T_{00}$ constant, the general solution of 
$$ t^{-1} \ddot{h} + t^{-2} \dot{h} -\frac{3}{4} t^{-3} = C $$
is of the form
$$ h(t) = \frac{C}{9} t^3 + B + A \log(t) + \frac{3}{8} \log(t)^2, $$
for arbitrary integration constants $A$ and $B$. 
With $C=(12 \pi^2 T_{00})/(96 f_2)$, this gives \eqref{wavesLamSol1}.

In the inflationary epoch where one has $a(t)=e^{\alpha t}$, for some $\alpha>0$, 
one can write the equation \eqref{wavesLambda2} in the form
$$ \ddot{h} +2 \alpha \dot{h} -3 \alpha^2= e^{2\alpha t} \frac{12 \pi^2 T_{00}}{96 f_2}. $$
Again assuming $T_{00}$ constant, the general solution of
$$  \ddot{h} +2 \alpha \dot{h} -3 \alpha^2= C e^{2\alpha t} $$
with a constant $C$ is of the form
$$ h(t)= \frac{3\alpha}{2} t + \frac{C}{8 \alpha^2} e^{2\alpha t} - \frac{A}{2\alpha} e^{-2\alpha t} + B, $$
for arbitrary integration constants $A$ and $B$. With $C=(12 \pi^2 T_{00})/(96 f_2)$ as above, 
this gives \eqref{wavesLamSol2}.
\endproof

The behavior of the solutions \eqref{wavesLamSol1} and \eqref{wavesLamSol2} should be compared with the analogous equations in the ordinary case where of the equation \eqref{friedeq} with $\kappa_0^2 = 8\pi G\, T_{00}$ independent of the energy scale $\Lambda$ and equal to the ordinary Newton constant. In this case, one obtains, in the radiation dominated case with $a(t)=t^{1/2}$ the equation
$$ \ddot{h} + t^{-1} \dot{h} -\frac{3}{4} t^{-2} = 8\pi G\, T_{00}, $$
which has general solution
$$ h(t) = 2\pi G T_{00} \, t^2 + B + A \log(t) + \frac{3}{8} \log(t)^2, $$
which differs from \eqref{wavesLamSol1} for the presence of a quadratic instead of cubic term. 
Similarly, in the case of the inflationary epoch, where one has $a(t)=e^{\alpha t}$, in the ordinary case one has the equation
$$  \ddot{h} +2 \alpha \dot{h} -3 \alpha^2=  8\pi G\, T_{00}, $$
which has general solution
$$ (\frac{4 \pi G T_{00}}{\alpha} + \frac{3\alpha}{2}) \, t\, + \frac{A}{2\alpha} e^{-2\alpha t} + B, $$
which differs from what we have in \eqref{wavesLamSol2} by the presence of an additional linear 
term instead of an exponential term. 

Similar examples with different forms of the factor $a(t)$ in the metric \eqref{atmetric}
can easily be derived in the same way. Choices of the parameter $f_2$ for which the
term $\fc(\Lambda)$ cannot be neglected will give rise to varying behaviors of the equations
both in the radiation dominated era and during inflation. However, in those cases the equations cannot be integrated exactly so we cannot exhibit explicit solutions.

\subsection{The $\lambda_0$-ansatz and the Higgs}\label{ansatzSec}

We mention here briefly another consequence of the {\em ansatz} \eqref{lambdaRun} on
the running of the coefficient $\lambda_0(\Lambda)$ between the electroweak and
the unification scales. This running is different from the one used in \cite{CCM}
to derive the Higgs mass estimate. In fact, the RGE equations themselves are
different, since in \cite{CCM} one only considers the RGE for the minimal standard
model and the boundary conditions at unifications are also significantly different 
from the ones used in \cite{AKLRS} that we use here (see the second appendix
for a more detailed discussion of the boundary conditions in \cite{CCM} and 
\cite{AKLRS}). Also in \cite{CCM} the coefficient $\lambda_0$ is assumed to
run as the $\lambda$ in the minimal standard model, with only the boundary
condition at unification relating it to the values of $\fa$ and $\fb$ (which are in turn
different from the values at unification according to the RGEs of \cite{AKLRS}).
Thus, one can check how adopting the {\em ansatz} \eqref{lambdaRun} for
$\lambda_0(\Lambda)$ together with the boundary conditions of \cite{AKLRS}
for $\lambda$, $\fa$ and $\fb$ affects the estimate for the Higgs mass. In \cite{CCM}
one obtains a heavy Higgs at around 170 GeV, by the value
$$ \sqrt{2\lambda} \frac{2 M}{g} \sim 170\, {\rm GeV}, $$
where $\lambda$ is the low energy limit of the RGE flow in the minimal Standard Model 
for the coefficient $\lambda_0$ and where $2M/g \simeq 246$ GeV is the Higgs vacuum. 
If we replace the running used in \cite{CCM} with the running 
of $\lambda_0(\Lambda)$ of  \eqref{lambdaRun} with the boundary condition 
of \cite{AKLRS}, the same estimate would deliver a much lower value
$$ \sqrt{2\lambda_0(\Lambda_{ew}) } \frac{2 M}{g} =
\sqrt{ \frac{2 \lambda(\Lambda_{ew}) \pi^2 \fb(\Lambda_{ew})}{f_0 \fa^2(\Lambda_{ew})} }
\frac{2 M}{g} \sim 158\, {\rm GeV}. $$
This looks potentially interesting in view of the fact that the projected window of
exclusion for the Higgs mass in \cite{Tevatron} starts at 158 GeV (see also \cite{LHCphys}).
This gives only a first possible indication that a more detailed analysis of the RGEs for
the standard model with Majorana mass terms, as in \cite{AKLRS}, and a
careful discussion of the boundary conditions at unification and of the running
of the coefficients in the asymptotic expansion of the spectral action may yield
a wider spectrum of possible behaviors for the Higgs field within these
noncommutative geometry models. This topic deserves more careful consideration
that is beyond the main focus of the present paper.

\subsection{Antigravity in the early universe}\label{antigravSec}

Cosmological model exhibiting a sign change in the effective Newton constant in the
early universe, due to the interactions of gravity and matter, were studied for instance
in \cite{Linde}, \cite{Pollock} or \cite{Zee}. Those models of antigravity are based on
the presence of a non-minimal conformal coupling of gravity to another field, with a
Lagrangian of the form
$$ \cL = \int \frac{-1}{16 \pi G} R \sqrt{g} + \frac{1}{12} \int R \varphi^2 \sqrt{g} + \cL(\varphi, A, \psi), $$
where the last term contains the kinetic and potential terms for the field $\varphi$ and its
interactions with other fields $A$, $\psi$. The conformal coupling of $R$ and $\varphi$ 
gives rise to an effective gravitational constant of the form
$$ G_{{\rm eff}}^{-1} = G^{-1} - \frac{4}{3} \pi \varphi^2 . $$
where $\varphi$ is treated as a constant, which is estimated in \cite{Linde} in terms of
the excess of neutrino over antineutrino density, the quartic interaction coefficient $\lambda$ of
$\varphi$ and the Weinberg angle. A decrease in $G_{{\rm eff}}$ produces a corresponding increase
in the Planck density (see, however, the criticism to this model discussed in \cite{Pollock}).
Antigravity sectors with negative effective gravitational constant as in \cite{Linde} were recently 
considered within various   approaches
based on extra dimensions and brane world models, see for instance
 \cite{GZBR}, \cite{GRS}, or from the point of view of moduli in
 heterotic superstring theory, as in \cite{Pollock2}.

Within the noncommutative geometry model it is also possible
to find scenarios where one has antigravity in the early universe. 
The same mechanism proposed in \cite{Linde} can be reproduced within 
the NCG model, due to the presence of the conformal coupling of the Higgs field
to gravity. However, there is another, independent mechanism that can also 
produce a sign change of the Newton constant between the
unification and electroweak cosmological phase transitions and which is only
due to the running of the effective Newton constant with the RGE equations of
the particle physics content of the model.  In fact, there are choices of the
parameter $f_2$ of the model for which the effective gravitational constant
undergoes a sign change. 

An example of this behavior is obtained if one chooses the value of the effective 
gravitational constant $G_{\rm eff}$ to be equal to the Newton constant at 
unification scale, $G_{\rm eff}(\Lambda_{unif},f_2)=G$,  and then runs it down 
with the RGE equations. 
Notice that the assumption $G_{\rm eff}(\Lambda_{unif},f_2)=G$ is the same
that was proposed in \cite{CCM}, but due to the different RGE analysis considered
here, the behavior we describe now is different from the one projected in \cite{CCM}. 
With the estimate used in Lemma 5.2 of \cite{CCM} for the Majorana mass terms at unification
(under the assumptions of flat space and negligible Higgs vacuum expectation) setting
the effective gravitational constant of the model equal to the Newton constant 
at unification energy only requires $f_2$ of the order of at most $f_2 \simeq 10^2$,
while using the boundary conditions of \cite{AKLRS} one find a larger value of $f_2$. 
In fact, we see that setting 
\begin{equation}\label{f2unif}
f_2 \simeq 18291.3 
\end{equation}
gives $G_{{\rm eff}}(\Lambda_{unif}) =G=(1.22086 \times 10^{19})^{-2}$.

\begin{figure}
\includegraphics[scale=0.6]{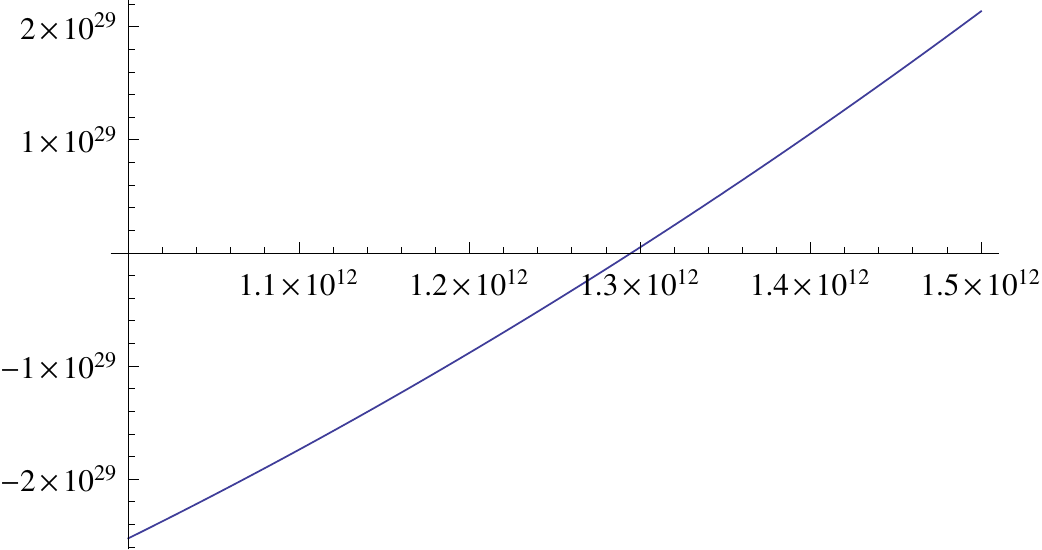}
\caption{An example of transition to negative gravity in the running of $G_{\rm eff}^{-1}(\Lambda)$ 
with  $G_{\rm eff}(\Lambda_{unif},f_2)=G$.
\label{GunifFig}}
\end{figure}

One sees then, as in Figure \ref{GunifFig}
that the resulting $G_{\rm eff}(\Lambda, f_2)$, for this choice of $f_2$
has a sign change at around $1.3\times 10^{12}$ GeV. Thus, with these boundary 
conditions one finds an example of a regime of negative gravity in the early universe. 
Other possible choices of $f_2$ lead to similar examples.

\subsection{The conformal gravity regime}\label{ConfGravSec}

In this scenario, it is especially interesting to see what happens near the energy scale of
$1.3\times 10^{12}$ GeV, where $G_{{\rm eff}}^{-1}$ vanishes, as in Figure \ref{GunifFig}.
This gives another example of an emergent conformal gravity regime at a phase
transition (here the change from positive to negative gravity) of the system.

\begin{prop}\label{confgravprop}
Let the parameter $f_2$ be chosen so that the inverse effective gravitational constant
$G_{\rm eff}^{-1}(\Lambda)$ has a zero at some $\Lambda=\Lambda_0$.
Assume the vanishing of the topological term and suppose that the parameter $f_4$
is chosen so that the effective cosmological constant also vanishes at $\Lambda_0$.
Then near $\Lambda_0$ and for $|H|^2$ sufficiently small the dynamics of 
\eqref{SAvarchange} is dominated by pure conformal gravity.
\end{prop}

\proof
At the singularity $\Lambda_{sing}$ for $G_{{\rm eff}}$, assuming the vanishing of the topological term, 
the terms that remain in the bosonic part of the action are the cosmological term, the
conformal gravity term with the Weyl curvature tensor, and the Higgs and gauge bosons
terms.  If the Higgs field is sufficiently near the $H=0$ vacuum, and the parameter $f_4$ is chosen
so that the cosmological term also vanishes at the same scale $\Lambda_{sing}$, 
one finds that what remains of the bosonic action is just the conformal gravity action
\begin{equation}\label{StermsGsing}
S(\Lambda_{sing}) = \alpha_0 \int C_{\mu\nu\rho\sigma} C^{\mu\nu\rho\sigma} \sqrt{g} d^4 x
\end{equation}
with $\alpha_0 =-3 f_0/(10\pi^2)\simeq -0.25916$, and 
with the additional weakly coupled term of the gauge bosons. Thus, in this scenario, when
running down the coefficients of the bosonic spectral action from unification scale towards
the electroweak scale, a singularity of $G_{{\rm eff}}$ occurs at an intermediate scale of
$1.3 \times 10^{12}$ GeV. At the singularity, if the cosmological term also vanishes and the Higgs contribution is sufficiently small, the model becomes dominated by a conformal gravity action \eqref{StermsGsing}.

In this regime the equations of motion will then be of the form
\begin{equation}\label{Ceqns}
2 C^{\mu\lambda\nu\kappa}_{;\lambda,;\kappa} - C^{\mu\lambda\nu\kappa} R_{\lambda\kappa}
= - \frac{1}{4\alpha_0} T^{\mu\nu}
\end{equation}
as in (188) of \cite{Mann}.
Thus one has in this regime pure conformal gravity with an effective gravitational constant
of  see \S 8.7 of \cite{Mann}.
\endproof

The emergence of a conformal gravity regime was also observed in \cite{NeSa}, though not
in terms of renormalization group analysis, but near the special value of the Higgs field
$|H| \to \sqrt{6}/\kappa_0$.

\subsection{Gravity balls}\label{BallsSec}

We now analyze more carefully the second mechanism that produces negative gravity
besides the running of $G_{\rm eff}(\Lambda)$, namely the non-minimal conformal
coupling to the Higgs field. This will reproduce in this model a scenario similar to
that of \cite{Linde}.

Let us assume for simplicity that the parameter $f_2$ is chosen so that
$G_{\rm eff}(\Lambda,f_2)>0$ for all $\Lambda_{ew} \leq \Lambda \leq 
\Lambda_{unif}$. We show that, even in this case, it is possible to have
regions of negative gravity, due to the coupling to the Higgs. 
These behave like the gravity
balls and non-topological solitons of \cite{Loh}, \cite{SaLo}, but with a
more elaborate behavior coming from the fact that the underlying 
gravitational constant is also changing with $\Lambda$ according to
the RGE flow.

\begin{prop}\label{antigravH}
Let $f_2$ be assigned so that the effective gravitational constant satisfies
$G_{\rm eff}(\Lambda,f_2)>0$ for all $\Lambda_{ew} \leq \Lambda \leq 
\Lambda_{unif}$. Then negative gravity regions, with $|H|$ near $|H|^2\sim \mu_0^2/(2\lambda_0)$,
arise in the range of energies $\Lambda$ such that
\begin{equation}\label{rangeantigrav}
\ell_H(\Lambda,f_2) > \ell_G(\Lambda,f_2),
\end{equation}
for 
\begin{equation}\label{vHiggsRun}
\ell_H(\Lambda,f_2) =  \frac{(2 f_2 \Lambda^2 \fa(\Lambda) - f_0 \fe(\Lambda) ) \fa(\Lambda)}{\pi^2 \lambda(\Lambda) \fb(\Lambda)}
\end{equation}
and
\begin{equation}\label{ellGeffRun}
\ell_G(\Lambda,f_2)= \frac{192 f_2 \Lambda^2 -2 f_0 \fc(\Lambda)}{4\pi^2}.
\end{equation}
\end{prop}

\proof
The presence of the conformal coupling term 
$$ - \frac{1}{12} \int R | H |^2 \sqrt{g} d^4 x $$
in the normalized asymptotic formula for the spectral action \eqref{SAvarchange}
means that, in regions with nearly constant $|H|^2$, the effective
gravitational constant of the model is further modified to give
\begin{equation}\label{GeffH}
G_{{\rm eff},H} = \frac{G_{{\rm eff}}}{1 - \frac{4 \pi}{3} G_{{\rm eff}} |H|^2}. 
\end{equation}
This is the same mechanism used in \cite{Linde} for negative
gravity models. This means that, assuming that $G_{{\rm eff}}(\Lambda)>0$ for
all $\Lambda_{ew} \leq \Lambda \leq \Lambda_{unif}$, one will have
$$ \left\{ \begin{array}{ll}
G_{{\rm eff},H} <0  & \text{for } |H|^2 > \displaystyle{\frac{3}{4\pi G_{{\rm eff}}(\Lambda)}}, \\[4mm]
G_{{\rm eff},H} > 0 & \text{for } |H|^2 < \displaystyle{\frac{3}{4\pi G_{{\rm eff}}(\Lambda)}}.
\end{array}\right.
$$ 
This means, for instance, that in the presence of an unstable equilibrium at
$|H|=0$ and a stable equilibrium at $|H|^2 = v^2$ satisfying 
$v^2 > \frac{3}{4\pi G_{{\rm eff}}(\Lambda)}$,
one can have gravity balls near zeros of the field $|H|^2$, where gravity behaves
in the usual attractive way, inside a larger scale negative gravity corresponding 
to the true equilibrium $|H|^2 =v^2$.

In the action \eqref{SAvarchange}, the Higgs field has a quartic potential given by
$$ -\mu_0^2 \int |H|^2 \sqrt{g} d^4x + \lambda_0 \int |H|^4 \sqrt{g} d^4 x, $$
which has a minimum at $\mu_0^2/(2\lambda_0)$. Thus, to identify the negative
gravity regime we need to compare the running of \eqref{vHiggsRun},
$$ \ell_H(\Lambda,f_2):= \frac{\mu_0^2}{2\lambda_0} (\Lambda)= \frac{2 \frac{f_2 \Lambda^2}{f_0} - \frac{\fe(\Lambda)}{\fa(\Lambda)} }{\lambda(\Lambda) \frac{\pi^2 \fb(\Lambda)}{f_0 \fa^2(\Lambda)}}
= \frac{(2 f_2 \Lambda^2 \fa(\Lambda) - f_0 \fe(\Lambda) ) \fa(\Lambda)}{\pi^2 \lambda(\Lambda) \fb(\Lambda)}, $$
where we again used the {\em ansatz} \eqref{lambdaRun} on
the running of the coefficient $\lambda_0(\Lambda)$, and the running of the function \eqref{ellGeffRun},
$$ \ell_G(\Lambda,f_2):= \frac{3}{4\pi G_{{\rm eff}}(\Lambda)} = \frac{3}{4\pi} \frac{192 f_2 \Lambda^2 -2 f_0 \fc(\Lambda)}{ 3\pi} = \frac{192 f_2 \Lambda^2 -2 f_0 \fc(\Lambda)}{4\pi^2}. $$
Thus, for different possible values of the parameter $f_2$, a negative gravity regime 
$G_{{\rm eff},H} <0$ and gravitational balls are possible in the range where
$\ell_H(\Lambda,f_2) > \ell_G(\Lambda,f_2)$. 
\endproof

\begin{figure}
\includegraphics[scale=0.8]{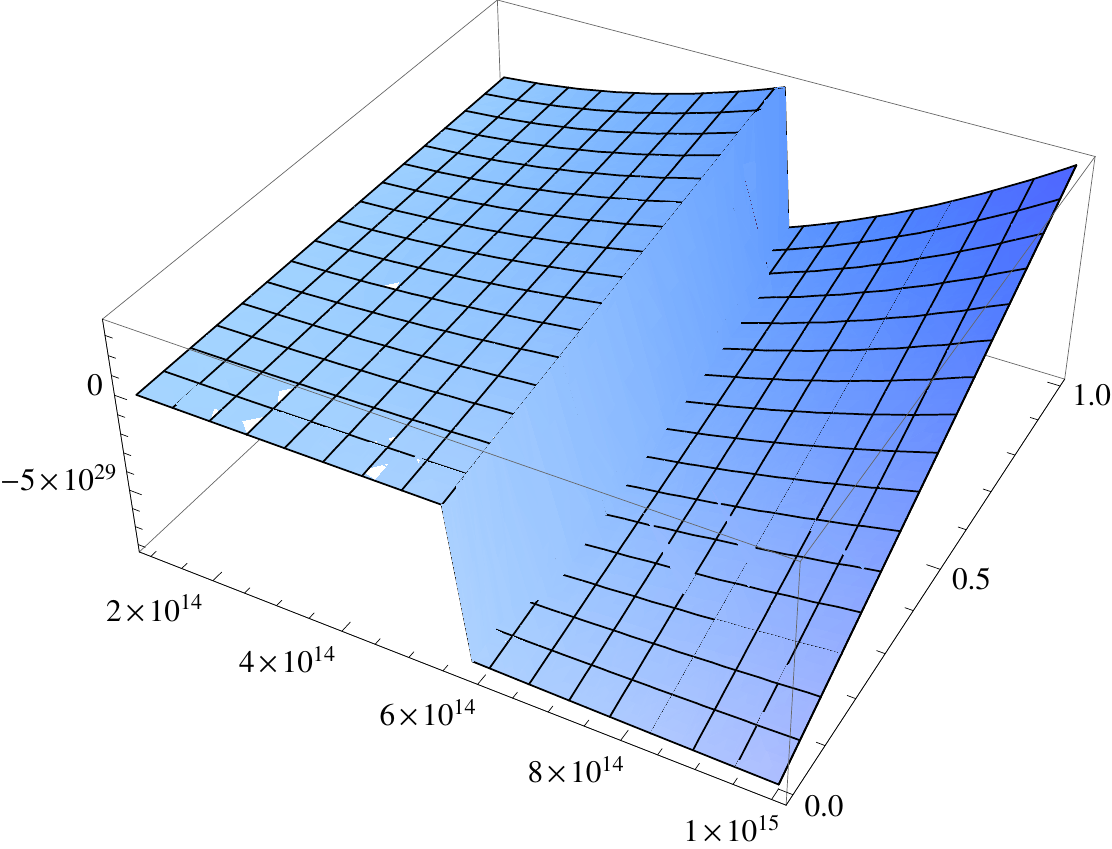}
\caption{The surface $\ell_H(\Lambda,f_2)$ in the range 
$10^{14}\leq \Lambda \leq 10^{15}$ GeV, around the top see-saw scale,
for $10^{-4} \leq f_2 \leq 1$. \label{ellHFig}}
\end{figure}

\begin{figure}
\includegraphics[scale=0.8]{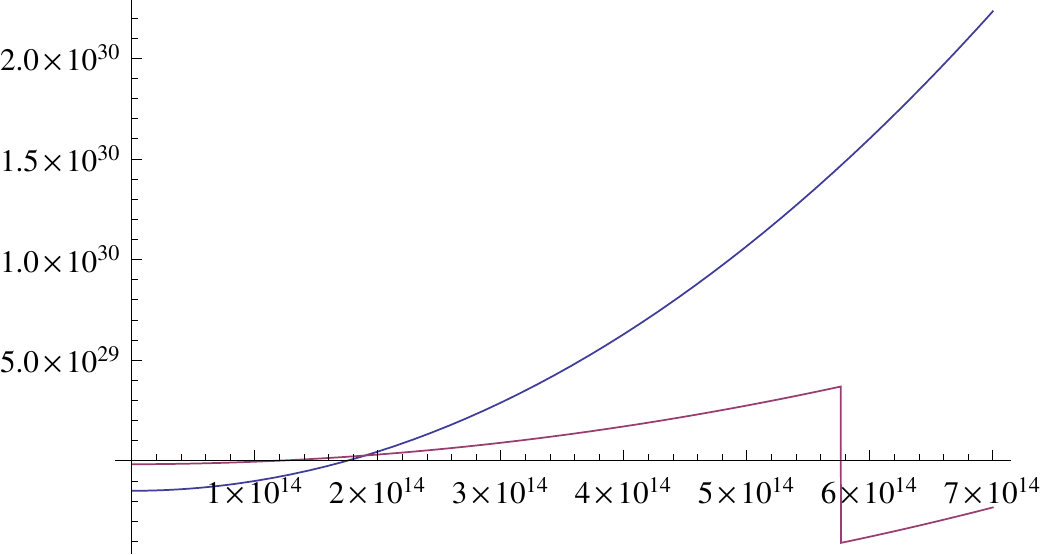}
\caption{Transition to a negative gravity regime where $\ell_H(\Lambda,f_2) > \ell_G(\Lambda,f_2)$
in the region $10^{12} \leq \Lambda \leq 10^{15}$ with $f_2=1$. \label{transneggravFig}}
\end{figure}

An example of such a transition to a negative gravity region is illustrated in Figure
\ref{transneggravFig}, where the surface $\ell_H(\Lambda,f_2)$ near the top see-saw
scale behaves as in Figure \ref{ellHFig}.

\subsection{Primordial black holes}\label{PBHsec}

The possibility of primordial black holes (PBHs) in the early universe
was originally suggested by Zeldovich and Novikov in the late 1960s 
(see \cite{Novikov} for a survey). They originate from the collapse of 
overdense regions, as well as from other mechanisms such as
phase transitions in the early universe, cosmic loops and strings, or
inflationary reheating.
The consequences on primordial black holes of the running of the gravitational
constant in the early universe have been analyzed, for example, in \cite{Carr}. 
In models where the gravitational constant in the early universe may be different
from the value it has in the modern universe, PBHs, whose existence is conjectured
but has not been presently confirmed, are seen as a possible source of 
information about the changing gravitational constant. In fact, the mass loss rate 
due to evaporation via Hawking radiation and the Hawking 
temperature of primordial black holes depends on the value of the Newton
constant so that the evolution of such black holes depends on the change
in the Newton constant. The evaporation of primordial black holes is
often proposed as a mechanism underlying $\gamma$-ray bursts (see 
for instance \cite{BeKoKo}, \cite{Buga}). An especially interesting question regarging
PBHs is that of {\em gravitational memory}, as described in \cite{Barrow}.
It is usually assumed that two possible scenarios for the evolution of
primordial black holes can happen: one where the evolution follows
the changing gravitational constant and one with the possibility of
``gravitational memory", namely where the evolution of the PBH is
determined by an effective gravitational constant different from the one of
surrounding space. This latter phenomenon can arise if the model has
the possibility of having regions, as in the case of the gravity balls discussed 
in \S \ref{BallsSec} above, where the effective gravitational constant has a 
value different from the one of surrounding space.

We discuss here the effect on primordial black holes of the running
of the effective gravitational constant in the noncommutative geometry model.
For primordial
black holes that formed at a time in the early universe when the gravitational constant
$G_{{\rm eff}}(t)$ was different from the one of the modern universe and whose evolution
in time reflects the corresponding evolution of the gravitational constant, the
black hole ``adjusts its size" to the changing $G_{{\rm eff}}(t)$ according to the equation
\begin{equation}\label{MasstG}
\frac{d \cM(t)}{dt} \sim - (G_{{\rm eff}}(t) \cM(t))^{-2} ,
\end{equation}
where $\cM$ is the mass of the primordial black hole. Correspondingly the
temperature varies with the changing gravitational constant as
\begin{equation}\label{Tblackhole}
T = (8\pi G_{{\rm eff}}(t) \cM(t))^{-1}.
\end{equation}
In the case with gravitational memory, one can have a black hole that
evolves according to a gravitational constant that is different to the one
of the surrounding space. In our setting, given the scenario described in
\S \ref{BallsSec}, this second case occurs with the equation \eqref{MasstG}
replaced by 
\begin{equation}\label{MasstGH}
\frac{d \cM(t)}{dt} \sim - (G_{{\rm eff},H}(t) \cM(t))^{-2} ,
\end{equation}
with $G_{{\rm eff},H}$ as in \eqref{GeffH}.

\begin{prop}\label{PBHprop}
In the radiation dominated era, for a cosmology with metric tensor of the form \eqref{atmetric},
the evaporation of primordial black holes by Hawking radiation in the NCG model
is given by
\begin{equation}\label{PBHevap}
 \cM(\Lambda,f_2) =\sqrt[3]{\cM^3(\Lambda_{in}) - \frac{2}{3\pi^2} \int_\Lambda^{\Lambda_{in}}
\frac{(192 f_2 x^2 - 2 f_0 \fc(x) )^2}{x^3}\, dx },
\end{equation}
in the case of PBHs without gravitational memory, while PBHs with gravitational memory
evaporate according to 
\begin{equation}\label{PBHevapGMem}
 \cM(\Lambda,f_2) =\sqrt[3]{\cM^3(\Lambda_{in}) - \frac{2}{3\pi^2} \int_\Lambda^{\Lambda_{in}}
\frac{(1-\frac{4\pi}{3} G_{\rm eff}(x)|H|^2)^2}{x^3 G_{\rm eff}(x)^2}\, dx },
\end{equation}
where
$$ G_{{\rm eff},H}^{-1}(\Lambda)= \frac{(1-\frac{4\pi}{3} G_{\rm eff}(\Lambda)|H|^2)}{G_{\rm eff}(\Lambda)} = $$ $$ \frac{192 f_2 \Lambda^2 - 2 f_0 \fc(\Lambda)}{3\pi} - \frac{4}{3\pi} \frac{(2 f_2 \Lambda^2 \fa(\Lambda) - f_0 \fe(\Lambda) )\fa(\Lambda)}{\lambda(\Lambda)\fb(\Lambda)}, $$
for $|H|^2 \sim \mu_0^2/(2\lambda_0)$. Here $\Lambda_{in}$ is the energy scale at which the
radiation dominated phase begins, that is, where one starts to have $a(t)=t^{1/2}$.
\end{prop}

\proof
For a metric of the form \eqref{atmetric}, in the radiation dominated era one has $a(t)\sim t^{1/2}$,
hence the energy--time change of variables $\Lambda = 1/a(t)$ is of the form $\Lambda = t^{-1/2}$.
Thus, the equation \eqref{MasstG} can be rewritten in the variable $\Lambda$ in
the form
\begin{equation}\label{MassLambda}
\frac{d \cM(\Lambda)}{d \Lambda} = \frac{2}{\Lambda^3 (G_{{\rm eff}}(\Lambda) \cM(\Lambda))^2},
\end{equation}
since $dt/d\Lambda =-2 \Lambda^{-3}$. 
We look at the case of a primordial black hole that has a given mass $\cM(\Lambda_{in})$ at the 
initial $\Lambda_{in}$, and we look at the evolution of
$\cM(\Lambda)$ between $\Lambda_{in}$ and electroweak scale.
The equation \eqref{MassLambda}, written in the form
$$ \cM(\Lambda)^2 \, d \cM(\Lambda) = 2 \frac{d \Lambda}{\Lambda^3 G_{\rm eff}^2(\Lambda)} $$
gives \eqref{PBHevap}. The case with gravitational memory is obtained similarly by replacing
the effective gravitational constant $G_{\rm eff}(\Lambda)$ with the one locally modified by the
interaction with the Higgs field, $G_{{\rm eff},H}(\Lambda)$.
\endproof

The expression for the evaporation by Hawking radiation simplifies in
the cases where $f_2$ is sufficiently large that the term $192 f_2 \Lambda^2$
dominates over $2 f_0 \fc(\Lambda)$.
This is the case, for example, when $G_{\rm eff}(\Lambda_{ew})=G$
and $f_2$ is chosen accordingly. 

\begin{cor}\label{BPHprop2}
In the case where $f_2$ is sufficiently large that the term $2 f_0 \fc(\Lambda)$
is negligible with respect to $192 f_2 \Lambda^2$ for all $\Lambda_{ew} \leq \Lambda \leq
\Lambda_{in}$, the evaporation law for a primordial black hole that forms at $\Lambda_{in}$
gives a bound on its mass by $(\Lambda_{in} 64\sqrt{6} f_2/\pi )^{2/3}$.
\end{cor}

\proof Under the assumption that $192 f_2 \Lambda^2$
dominates over $2 f_0 \fc(\Lambda)$, the right hand side of \eqref{MassLambda}
can be approximated by the dominant term of $G_{\rm eff}(\Lambda)$ which gives
\begin{equation}\label{MassLambdaAppr}
\cM^2 \, d\cM = 2 \left(\frac{64 f_2}{\pi}\right)^2 \, \Lambda \, d\Lambda.
\end{equation}
Let $\Lambda_0$ denote the scale at which the PBH evaporates. Then \eqref{MassLambdaAppr} 
gives 
$$ \frac{1}{3} \cM^3(\Lambda)=2 \left( \frac{(64 f_2)}{\pi} \right)^2 (\Lambda^2 -\Lambda_0^2). $$
This sets a bound to the mass of a primordial black hole formed at $\Lambda=\Lambda_{in}$ 
from the condition that
$$ \Lambda_0^2 = \Lambda_{in}^2 - \left(\frac{\pi}{(64 f_2 \sqrt{6})}\right)^2 \cM^3_{unif} , $$
which gives 
$$ \cM_{in} \leq \left(\frac{ \Lambda_{in}  64 f_2 \sqrt{6}}{\pi} \right)^{2/3}. $$
\endproof

We can analyze similarly the equation of the Hawking radiation for PBHs during 
the inflationary epoch when one has $a(t)=e^{\alpha t}$ for some $\alpha >0$.
We obtain the following result.

\begin{prop}\label{PBHpropInfl}
In the inflationary epoch, for a cosmology with metric tensor of the form \eqref{atmetric},
the evaporation of primordial black holes by Hawking radiation in the NCG model
is given by
\begin{equation}\label{PBHevapInfl}
 \cM(\Lambda,f_2) =\sqrt[3]{\cM^3(\Lambda_{in}) - \frac{1}{3\alpha \pi^2} \int_\Lambda^{\Lambda_{in}}
\frac{(192 f_2 x^2 - 2 f_0 \fc(x) )^2}{x}\, dx },
\end{equation}
in the case of PBHs without gravitational memory, while PBHs with gravitational memory
evaporate according to 
\begin{equation}\label{PBHevapGMemInfl}
 \cM(\Lambda,f_2) =\sqrt[3]{\cM^3(\Lambda_{in}) - \frac{1}{3\alpha\pi^2} \int_\Lambda^{\Lambda_{in}}
\frac{(1-\frac{4\pi}{3} G_{\rm eff}(x)|H|^2)^2}{x G_{\rm eff}(x)^2}\, dx }.
\end{equation}
where $\Lambda_{in}$ is the energy scale at which the inflationary behavior
$a(t)=e^{\alpha t}$ begins.
\end{prop}

\proof The argument is completely analogous to the previous case. Here the energy-time 
change of variables $\Lambda = 1/a(t)$ gives $t = - \alpha^{-1} \log\Lambda$. Thus, in the
energy variable the equation \eqref{MasstG} becomes
$$ \cM^2 \, d\cM = \frac{d\Lambda}{ \alpha \Lambda (G_{\rm eff}(\Lambda))^2} $$
for the case without gravitational memory, or the same equation with $G_{\rm eff}(\Lambda)$
replaced by $G_{{\rm eff}, H}(\Lambda)$ in the case with gravitational memory. This gives
\eqref{PBHevapInfl} and \eqref{PBHevapGMemInfl}.
\endproof

\subsection{Higgs based slow-roll inflation}\label{HiggsInflSec}

Recently, a mechanism for inflation within the minimal standard model physics
was proposed in \cite{dSHW}. It is based on the presence of a non-minimal
coupling of the Higgs field to gravity of the form 
$$ - \xi_0 \int  R\, |H|^2 \, \sqrt{g} \, d^4x $$
as we have in the asymptotic expansion of the spectral action in the
noncommutative geometry model, but where the value of $\xi_0$ is not
set equal to the conformal coupling $\xi_0=1/12$, but is subject to running 
with the RGE flow. In \cite{dSHW} the running of $\xi_0$ is governed by the
beta function given in \cite{BOS}, which in our notation we can write approximately as
\begin{equation}\label{xi0beta}
16 \pi^2 \beta_{\xi_0}(\Lambda)  = (-12 \xi_0(\Lambda) + 1)\,
F(Y_u,Y_d,Y_{\nu},Y_e,M,g_1,g_2,g_3,\lambda),
\end{equation}
where the function $F$ of the running parameters of the model is computed explicitly
in \cite{BOS}. In  \cite{dSHW} this running is only considered within the minimal
standard model, without the right handed neutrinos and Majorana mass terms $M$,
but their argument can be adapted to this extension of the standard model, since the
general derivation of the running of $\xi_0$ in \cite{BOS} applies in greater generality. 
In this variable $\xi_0$ scenario, the dimensionless quantity that governs inflation
is $\psi = \sqrt{\xi_0} |H| /m_P$, where $m_P$ is the reduced Planck mass, which in our
notation is $m_P^2 =1/\kappa_0^2$. The inflationary period corresponds in \cite{dSHW}
to the large values $\psi >> 1$, the end of the inflation to the values $\psi \sim 1$ and the
low energy regime to $\psi << 1$.

At present we do not know whether a modification of the noncommutative geometry
model that allows for a variable $\xi_0(\Lambda)$,
different from the conformal coupling $\xi_0=1/12$ is possible within the costraints
of the model, but we show here that, even in the case where $\xi_0$ is constant 
in $\Lambda$ and equal to the conformal $\xi_0=1/12$, the 
noncommutative geometry model still allows for a similar
inflation mechanism to occur, through the running of the effective gravitational constant.

\begin{prop}\label{VEprop}
A Higgs based slow roll inflation scenario arises in the NCG model with
parameter
$$ \psi(\Lambda)^2= \xi_0 \kappa_0^2(\Lambda) |H|^2 $$
and with potential
\begin{equation}\label{VEHform}
 V_E(H)= \frac{\lambda_0 |H|^4}{(1+\xi_0 \kappa_0^2 |H|^2)^2} . 
\end{equation}
\end{prop}

\proof
In the noncommutative geometry model, the coefficient $\kappa_0(\Lambda)$ is
running with $\Lambda$ according to
$$ \kappa_0^2(\Lambda) = \frac{12 \pi^2}{96 f_2 \Lambda^2 - f_0 \fc(\Lambda)}. $$
Then the parameter that controls inflation is given by
$$ \psi(\Lambda)^2= \xi_0(\Lambda) \kappa_0^2(\Lambda) |H|^2 =
 \frac{\pi^2}{96 f_2 \Lambda^2 - f_0 \fc(\Lambda)} \, |H|^2. $$

If one then proceeds as in \cite{dSHW}, one finds that 
In the Einstein metric $g^E_{\mu\nu} = f(H) g_{\mu\nu}$, for $f(H)=1+\xi_0 \kappa_0 |H|^2$
the Higgs potential becomes of the form \eqref{VEHform}.
In the range where $\psi >>1$ this approaches the constant function (constant in $H$ but not in $\Lambda$)
$$ V_E= \frac{\lambda_0(\Lambda)}{4 \xi_0^2(\Lambda) \kappa_0^4(\Lambda)} = 
 \frac{\lambda(\Lambda) \fb(\Lambda) (96 f_2 \Lambda^2 - f_0 \fc(\Lambda))^2}
{4 f_0 \fa^2(\Lambda) }, $$
where we used again the {\em ansatz} \eqref{lambdaRun} for the running of the
coefficient $\lambda_0(\Lambda)$, 
while at low values $\psi <<1$ the potential is well approximated by the usual quartic potential
$V_E(H)\sim \lambda_0 |H|^4$. 
\endproof

\begin{figure}
\includegraphics[scale=0.6]{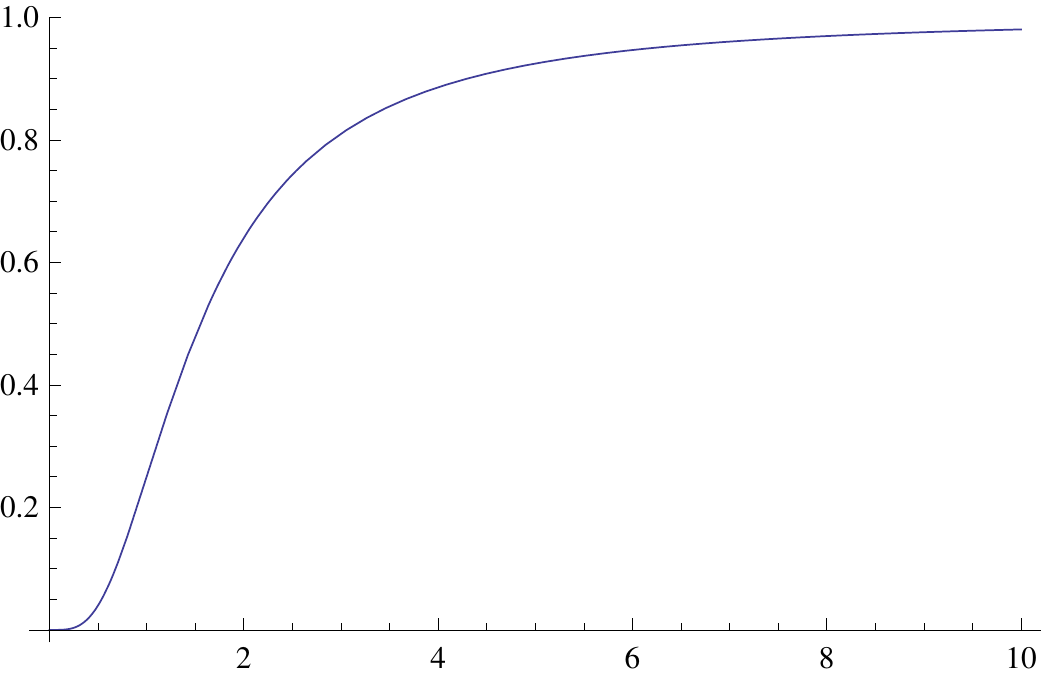}
\caption{Plot of $V_E(H)/V_E$ against $\psi$. \label{psiVFig}}
\end{figure}

\begin{figure}
\includegraphics[scale=0.5]{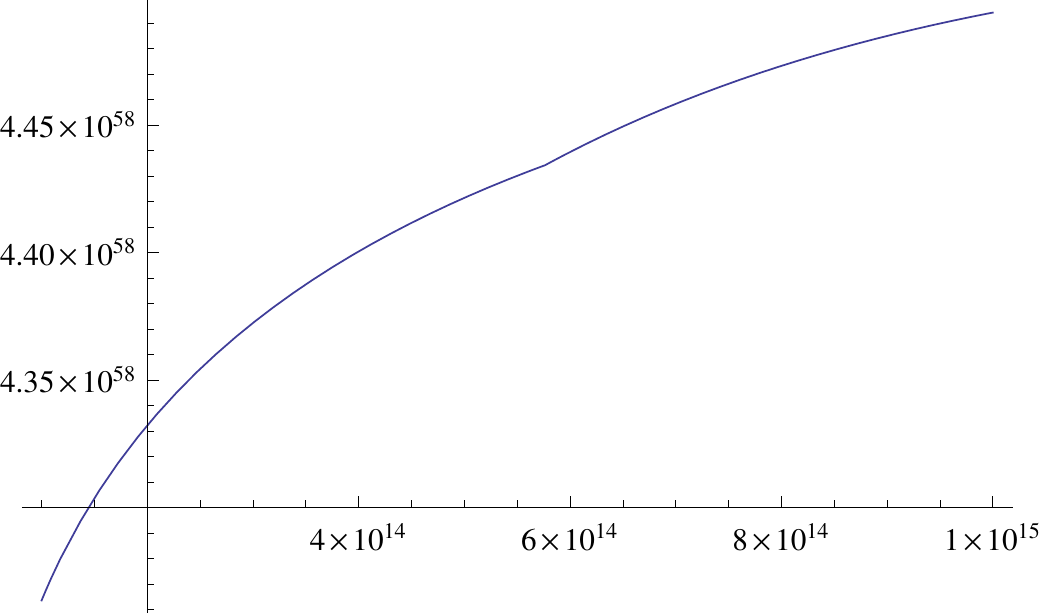}
\includegraphics[scale=0.5]{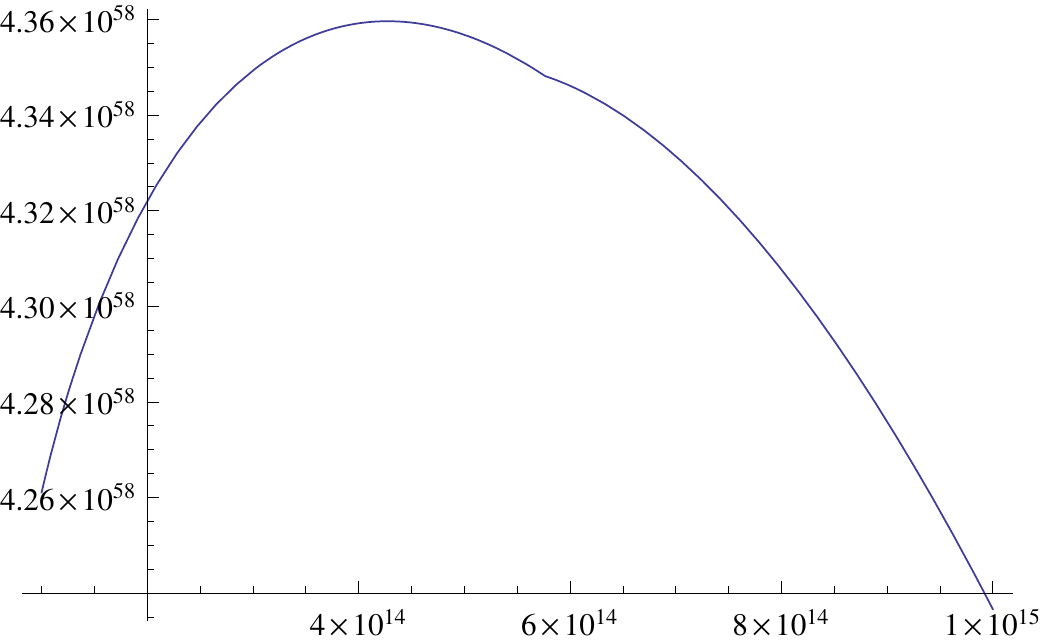}
\includegraphics[scale=0.5]{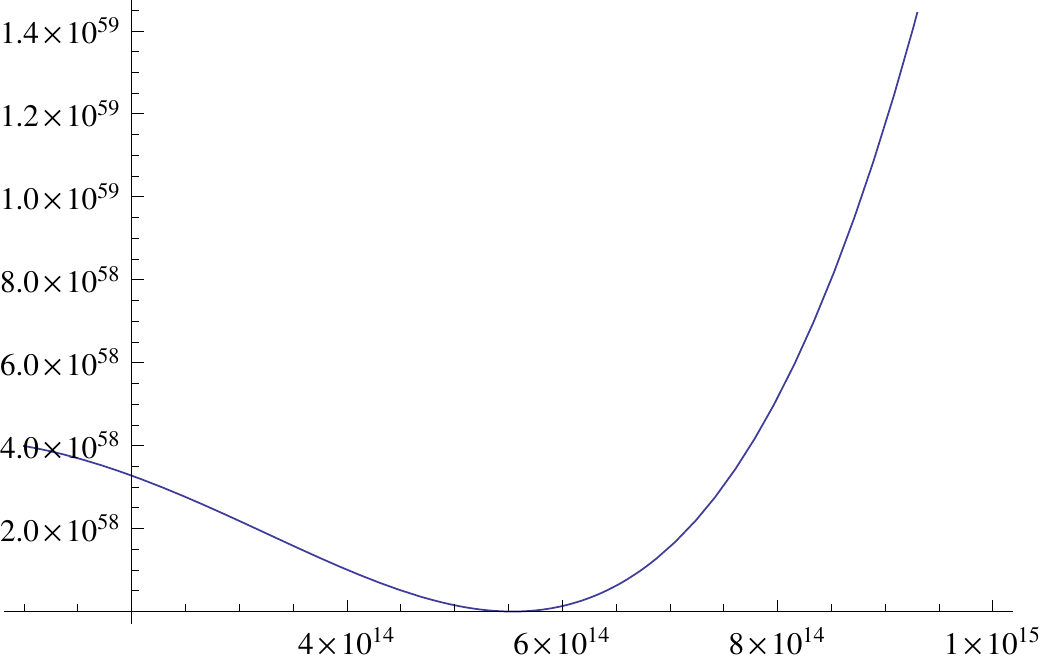}
\caption{The behavior of the asymptotic value $V_E$ of the potential $V_E(H)$, plotted above
for $f_2=10^{-4}$, $f_2=10^{-3}$, and $f_2=1/10$ and for $10^{14}\leq \Lambda \leq 10^{15}$  \label{VEplotFig}}
\end{figure}

The asymptotic value $V_E$ in turn depends on the energy scale $\Lambda$ 
and different behaviors are possible upon changing the values of the
parameter $f_2$ of the model as the examples in Figure \ref{VEplotFig}
illustrate. One can see the effect of the top see saw-scale on the running.

Notice that a potential of the form as in Figure \ref{psiVFig}, which is especially suitable for
slow roll inflationary models, also arises naturally in the NCG setting from
non-perturbative effects in the spectral action, as in \cite{CC3}. We plan to return
in future work to describe how the non-perturbative approach can be applied to
cosmological models beyond the very early universe. 

\subsection{Spectral index and tensor to scalar ratio}

In a slow roll inflation model, the first and second derivatives of the potential $V_E$
together with a change of variable in the field that brings the action into a canonical
form, determine the first and second slow-roll parameters $\epsilon$ and $\eta$,
see \cite{LidLy} \S 3.4 and also the analysis in \cite{dSHW}.
We analyze here the form of the slow-roll parameters in the noncommutative geometry model.

We write the potential $V_E$ of \eqref{VEHform} in the form
\begin{equation}\label{VEs}
 V_E(s)= \frac{\lambda_0 s^4}{(1+\xi_0 \kappa_0^2 s^2)^2} . 
\end{equation}
We will consider derivatives of this potential as a function of the variable
$s$.

We also introduce the expression 
\begin{equation}\label{Csnewfield}
C(s):= \frac{1}{2 (1+\xi_0 \kappa_0^2 s^2)} + \frac{3}{2\kappa_0^2} 
\frac{(2 \xi_0 \kappa_0^2 s)^2}{(1+\xi_0 \kappa_0^2 s^2)^2},
\end{equation}
which corresponds to the $(d\sigma/d\phi)^2$ of equation (4) of \cite{dSHW}
that gives the change of variable in the field that runs the inflation that puts
the action in a canonical form. 

The two slow-roll parameters are then given by the expressions
\begin{equation}\label{slowepsilon}
\epsilon(s) = \frac{1}{2 \kappa_0^2} \left( \frac{V_E^\prime(s)}{V_E(s)} \right)^2 C(s)^{-1},
\end{equation}
\begin{equation}\label{sloweta}
\eta(s) = \frac{1}{\kappa_0^2} \left( \frac{V_E^{\prime\prime}(s)}{V_E(s)} C(s)^{-1} -
\frac{V_E^\prime(s)}{V_E(s)} C(s)^{-3/2} \frac{d}{ds} C(s)^{1/2}\right) ,
\end{equation}
which correspond to the equations (6) and (7) of \cite{dSHW}.
We then have the following result.

\begin{prop}\label{slowrolls}
The first slow roll coefficient is of the form
\begin{equation}\label{espilonNCG}
\epsilon(s)=\frac{16 \kappa_0^2}{s^2 + \xi_0 \kappa_0^2 (1+ (\kappa_0^2)^2 ) s^4} 
\end{equation}
while the second slow-roll coefficient is given by
\begin{equation}\label{etaNCG}
\eta(s)= \frac{
8 (3 + \xi_0 \kappa_0^2 s^2 (1 - 2 \xi_0 \kappa_0^2 (s^2 + 12 \kappa_0^2 (-1 + \xi_0 \kappa_0^2 s^2))))}
{\kappa_0^2 (s + \xi_0 \kappa_0^2 (1 + (\kappa_0^2)^2) s^3)^2}.
\end{equation} 
\end{prop}

\proof The derivatives of the potential are
$$ V_E^\prime(s)= - \frac{4 \lambda_0 \xi_0 \kappa_0^2 s^5}{(1 + \xi_0 \kappa_0^2 s^2)^3}
+ \frac{4 \lambda_0 s^3}{(1 + \xi_0 \kappa_0^2 s^2)^2} $$
$$ V_E^{\prime\prime}(s)= \frac{24 \lambda_0 \xi_0^2 (\kappa_0^2)^2 s^6}
{(1 + \xi_0 \kappa_0^2 s^2)^4} - \frac{36 \lambda_0 \xi_0 \kappa_0^2 s^4}{(1 + \xi_0 \kappa_0^2 s^2)^3}
+ \frac{12 \lambda_0 s^2}{(1 + \xi_0 \kappa_0^2 s^2)^2}. $$
One also has
$$ C(s)^{-3/2} = \frac{ \sqrt{8}(1 + \xi_0 \kappa_0^2 s^2)^3}{\sqrt{
(1 + \xi_0 \kappa_0^2 (1 + 12 \xi_0 \kappa_0^2) s^2)^3} }$$
$$ \frac{d}{ds} C(s)^{1/2} = - \frac{\xi_0 \kappa_0^2 s (1 + \xi_0 \kappa_0^2 (s^2 + 12 \kappa_0^2
 (-1 + \xi_0 \kappa_0^2 s^2)))}{ (1 + \xi_0 \kappa_0^2 s^2)^2 \sqrt{ 2(
 1 + \xi_0 \kappa_0^2 (1 + 12 \xi_0 (\kappa_0^2)^2) s^2 } } . $$
One then computes $\epsilon(s)$ and $\eta(s)$ directly from \eqref{slowepsilon} and
\eqref{sloweta}. 
\endproof

Notice that both coefficients depend on the energy scale $\Lambda$ and on 
the parameters $f_2$ and $f_0$ through the single parameter 
$$ \kappa_0^2(\Lambda) = \frac{12 \pi}{96 f_2 \Lambda^2 -f_0 \fc(\Lambda)}, $$
while no dependence through $\lambda_0(\Lambda)$ remains and in our
model $\xi_0=1/12$.

The slow-roll coefficients provide expressions for the spectral index $n_s$ and
the tensor to scalar ratio $r$, which can be directly compared with cosmological data.
Thus, a more detailed analysis of the behavior of these as functions of $\Lambda$
and of the parameter $f_2$ of the model will give an exclusion curve for the
parameter. We will provide a more detailed analysis elsewhere, but for the
purpose of the present paper we derive the corresponding expression one
has in this model for the spectral index and the tensor to scalar ratio. 

\begin{prop}\label{nsr}
The spectral index is of the form
\begin{equation}\label{ns}
n_s = 1 + \frac{32 (216 + \kappa_0^2 (6 s^2 - \kappa_0^2 (432 + 12 \kappa_0^2 (2 + 3 (\kappa_0^2)^2) s^2 + (1 + (\kappa_0^2)^2) s^4)))}
{ \kappa_0^2 (12 s +  \kappa_0^2(1 + ( \kappa_0^2)^2) s^3)^2},
\end{equation}
while the tensor to scalar ratio if given by
\begin{equation}\label{rtensscal}
r =  \frac{256 \kappa_0^2}{s^2 + \frac{\kappa_0^2}{12} (1+ (\kappa_0^2)^2 ) s^4} .
\end{equation}
\end{prop}

\proof One knows that the spectral index and the tensor to scalar ratio are related
to the slow-roll coefficients by (see \cite{LidLy} \S 7.5.2 and \S 7.6 and \cite{dSHW})
\begin{equation}\label{nsandr}
n_s= 1- 6 \epsilon + 2 \eta, \ \ \ \text{ and } \ \ \  r = 16 \epsilon.
\end{equation}
The result then follows directly from Proposition \ref{slowrolls}.
\endproof

Once again, these parameters have a dependence on the choice of the free
parameter $f_2$ of the model and a scaling behavior with the energy $\Lambda$,
which depends on the running of $\kappa_0^2(\Lambda)$. Since the spectral
index and the tensor to scalar ratio are heavily constrained by cosmological
data from the WMAP combined with baryon acoustic oscillations and supernovae data,
this provides a way to impose realistic constraints on the parameter $f_2$ of the model
based on direct confrontation with data of cosmological observations. We will
provide a detailed analysis of the constraints imposed on the noncommutative
geometry model by the spectral index and tensor to scalar ratio in a separate paper.
Since the parameters $n_s$ and $r$ are also providing information on the gravitational
waves (see \cite{LidLy} \S 7.7) a more detailed analysis of their behavior and
dependence on the parameters of the model will give us a better understanding
of the effects on gravitational waves of the presence of noncommutativity
and may explain in a different way the amplification phenomena in the
propagation of gravitational waves that we observed in \S \ref{gravwaveSec}.

\subsection{Variable effective cosmological constant}\label{cosmoconstSec}

The relation between particle physics and the cosmological constant, through
the contribution of the quantum vacua of fields, is well known since the seminal
work of Zeldovich \cite{Zeldo}. The cosmological constant problem is the
question of reconciling a very large value predicted by particle physics with
a near zero value that conforms to the observations of cosmology.  Among the
proposed solutions to this problem are various models, starting with \cite{Zeldo}, 
with a varying effective cosmological constant, which would allow for a large 
cosmological constant in the very early universe, whose effect of negative pressure 
can overcome the attractive nature of gravity and result in accelerated
expansion, and then a decay of the cosmological constant to zero (see also
\cite{OvCo} for a more recent treatment of variable cosmological constant models). Often the
effective cosmological constant is produced via a non-minimal coupling
of gravity to another field, as in \cite{Dolgov}, similarly to what one
does in the case of an effective gravitational constant.

In the present model, one can recover the same mechanism of \cite{Dolgov}
via the non-minimal coupling to the Higgs field, but additionally one has a
running of the effective cosmological constant $\gamma_0(\Lambda)$ which
already by itself may produce the desired effect of decaying cosmological
constant. We illustrate in this section an example of how different choices
of the parameter $f_4$, for a fixed choice of $f_2$, generate different
possible decay behaviors of the cosmological constant. These can then
be combined with the effect produced by the non-minimal coupling with
the Higgs field, which behaves differently here than in the case originally
analyzed by \cite{Dolgov}. This still does not resolve the fine tuning
problem, of course, because we are trading the fine tuning of the
cosmological constant for the tuning of the parameters $f_2$ and $f_4$
of the model, but the fact that these parameters have a geometric
meaning in terms of the spectral action functional may suggest 
geometric constraints.

As we have seen in the proof of Proposition \ref{dominantprop} above, one can
impose the vanishing of the effective cosmological constant $\gamma_0$ 
at a given energy scale $\Lambda$ by fixing the parameter $f_4$
equal to the value given in \eqref{f4Lambda}. This means that,
when the effective gravitational constant of the model varies as in the
effective gravitational constant surface $G_{\rm eff}(\Lambda,f_2)$,
depending on the value assigned to the parameter $f_2$, there is
an associated surface that determines the value of the parameter $f_4$
that gives a vanishing effective cosmological constant. This is defined by the
equation
\begin{equation}\label{f4surface}
\pi^2 \gamma_0(\Lambda,f_2,f_4)= 48 f_4 \Lambda^4 - f_2 \Lambda^2 \fc(\Lambda) + \frac{1}{4} f_0 \fd(\Lambda) =0 .
\end{equation}
The solutions to this equation determine the surface
\begin{equation}\label{f4surf}
 f_4(\Lambda,f_2) = \frac{f_2 \Lambda^2 \fc(\Lambda)- \frac{1}{4} f_0 \fd(\Lambda)}{48 \Lambda^4}.
\end{equation}

To illustrate the different possible behaviors of the system when imposing the
vanishing of the effective cosmological constant at different possible energy
scales $\Lambda$, we look at the examples where one imposes a vanishing 
condition at one of the two ends 
of the interval of energies considered, that is, for $\Lambda=\Lambda_{ew}$ or for $\Lambda=\Lambda_{unif}$. For simplicity we also give here an explicit example for a fixed
assigned value of $f_2$, which first fixes the underlying $G_{\rm eff}(\Lambda)$.
we choose, as above, the particular example where $f_2$ is chosen to satisfy
$G_{\rm eff}(\Lambda_{ew})= G$.
Under these conditions we impose the vanishing of the effective cosmological
constant at either $\Lambda_{ew}$ or $\Lambda_{unif}$ through the
constraint
\begin{equation}\label{f4ew0}
\gamma_0(\Lambda_{ew},f_2,f_4)=0  \ \ \ \text{ or }  \ \ \  \gamma_0(\Lambda_{unif},f_2,f_4)=0
\end{equation}
which fixes the value of $f_4=f_4(\Lambda_{ew},f_2)$. Then the behavior at higher
energies of the effective cosmological constant is given by the surface
\begin{equation}\label{gammaf2Lambda}
\gamma_0(\Lambda,f_2,f_4(\Lambda_{ew},f_2)) \ \ \ \text{ or }  \ \ \ \gamma_0(\Lambda,f_2,f_4(\Lambda_{unif},f_2))
\end{equation} 
as a function of the energy $\Lambda$ and the parameter $f_2$. 

\begin{figure}
\includegraphics[scale=0.55]{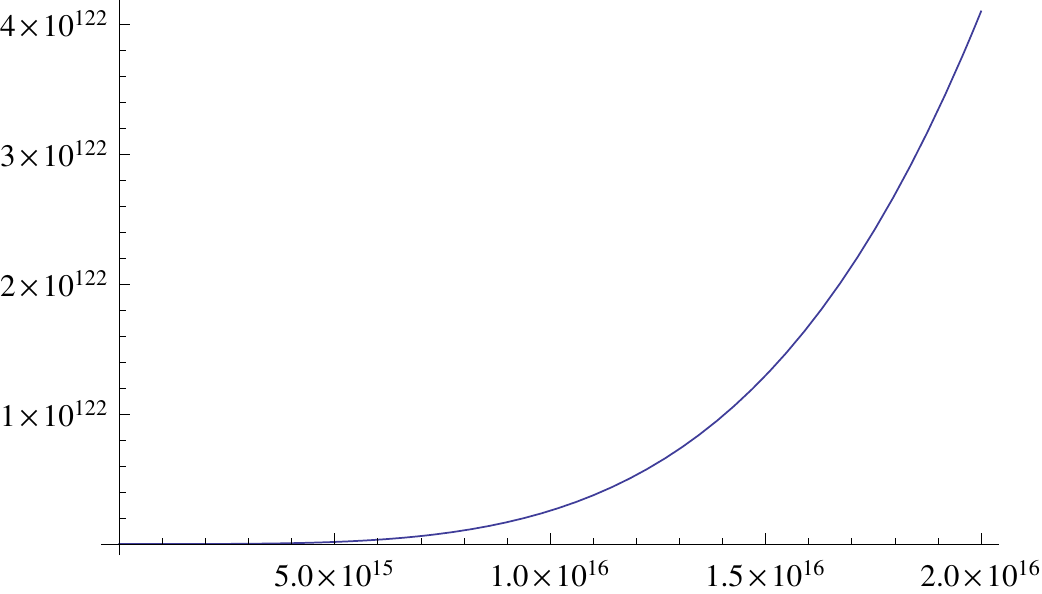}
\includegraphics[scale=0.55]{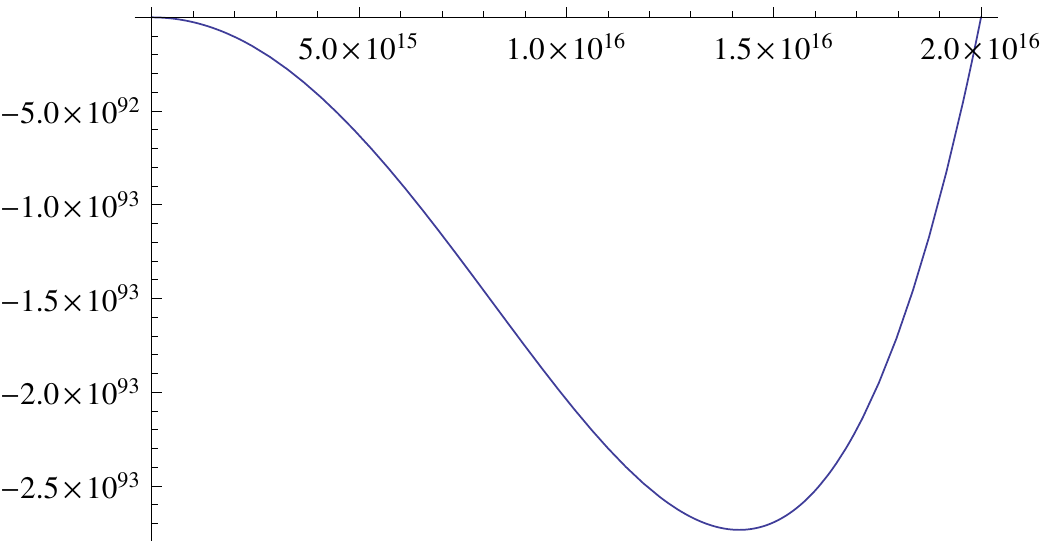}
\caption{Running of $\gamma_0(\Lambda)$ under the assumptions of vanishing
$\gamma_0(\Lambda_{ew})=0$ or $\gamma_0(\Lambda_{unif})=0$.\label{gamma0Fig}}
\end{figure}

We find two very different behaviors in these two examples, which have different
implications in terms of dark energy. Imposing the vanishing of $\gamma_0$
at $\Lambda_{ew}$ produces a very fast growth of $\gamma_0(\Lambda)$ at larger 
energies $\Lambda_{ew}\leq \Lambda \leq \Lambda_{unif}$, while imposing the 
vanishing at $\Lambda_{unif}$ produces a $\gamma_0(\Lambda)$ that quicky
decreases to a large negative value and then slowly decreases to become small
again near $\Lambda_{ew}$, as illustrated in the two examples of Figure \ref{gamma0Fig}.
A negative cosmological constant adds to the attractive nature of gravity hence it
counteracts other possible inflation mechanism that may be present in the model,
while a positive cosmological constant counteracts the attractive nature of gravity
and can determine expansion. Thus, we see in these two examples that different 
choices of the free parameter $f_4$ in the model can lead to very different qualitative 
behaviors of the effective cosmological constant. 

The range of different variable cosmological
constants can be further expanded by considering other possible values of the
parameter $f_4$ in the surface $\gamma_0(\Lambda,f_4,f_2)$ associated to a
chosen value of $f_2$.

Again, as we have seen in the case of the effective gravitational constant, we
have here two distinct mechanisms for vacuum-decay: the running 
$\gamma_0(\Lambda)$ as described above, depending
on the parameters $f_2$ and $f_4$ and on the renormalization group equations, 
and a further modification due to the non-minimal coupling with the Higgs field,
as in \cite{Dolgov}.

Namely, if one has an unstable and a stable equilibrium for $|H|^2$, one can
encounter for the cosmological constant the same type of phenomenon described 
for the effective gravitational constant by the gravity balls, where the value of 
$\gamma_{0,H}$ is equal to $\gamma_0$ at the unstable equilibrium $|H|^2=0$ 
on some large region, while outside that region it decays to the stable equilibrium $|H|^2 =v^2$
for which 
$$ \gamma_{0,H}(\Lambda)=\frac{\gamma_0(\Lambda)}{1- 16 \pi G_{\rm eff}(\Lambda) \xi_0 v^2}. $$
This behaves differently in the case of our system from the model of \cite{Dolgov}. 
In fact, here the non-minimal coupling is always the conformal one $\xi_0=1/12$, while
the gravitational constant is itself running, so we obtain the following.

\begin{prop}\label{gamma0Hprop}
The non-minimal conformal coupling of the Higgs field to gravity,
$$ -\xi_0 \int R\, |H|^2\, \sqrt{g}\, d^4x $$
changes the running of the effective cosmological constant to
\begin{equation}\label{effcosmoH}
 \gamma_{0,H}(\Lambda) =\frac{\gamma_0(\Lambda)}{1- 16 \pi G_{\rm eff}(\Lambda) \xi_0 |H|^2} .
\end{equation} 
This gives
$$ \gamma_{0,H}(\Lambda) = \frac{\gamma_0(\Lambda)}
{1- \frac{ 4\fa(\Lambda) (2f_2 \Lambda^2 \fa(\Lambda) - f_0 \fe(\Lambda) )}{\lambda(\Lambda)\fb(\Lambda) (192 f_2 \Lambda^2 - 2 f_0 \fc(\Lambda))}}, $$
for $|H|^2 \sim \mu_0^2/(2\lambda_0)$.
\end{prop}

\proof Arguing as in \cite{Dolgov} we obtain \eqref{effcosmoH}, which now 
depends on the running of both $\gamma_0(\Lambda)$ and $G_{\rm eff}(\Lambda)$, 
as well as on the Higgs field $|H|^2$. Assuming the latter to be nearly constant
with $|H|^2 \sim \mu_0^2/(2\lambda_0)$, we obtain
\begin{equation}\label{gamma0Grun}
\gamma_{0,H}(\Lambda)= \gamma_0(\Lambda) \frac{G_{{\rm eff},H}(\Lambda)}{G_{\rm eff}(\Lambda)}= \frac{\gamma_0(\Lambda)}{1 - \frac{4\pi}{3} G_{\rm eff}(\Lambda) \left(
\frac{2 f_2 \Lambda^2 \fa(\Lambda)- f_0 \fe(\Lambda) \fa (\Lambda)}{\pi^2 \lambda(\Lambda)
\fb(\Lambda)}\right) } ,
\end{equation}
where, as in the treatment of the variable effective gravitational constant, we have
$$ |H|^2\sim \frac{\mu_0^2}{2\lambda_0} =  
\frac{2 f_2 \Lambda^2 \fa(\Lambda)- f_0 \fe(\Lambda) \fa (\Lambda)}{\pi^2 \lambda(\Lambda)
\fb(\Lambda)} $$
using the {\em ansatz} \eqref{lambdaRun} for $\lambda_0(\Lambda)$, and 
$G_{{\rm eff},H}^{-1}(\Lambda)=G_{\rm eff}^{-1}(\Lambda) - \frac{4\pi}{3} |H|^2$.
\endproof

\begin{figure}
\includegraphics[scale=0.55]{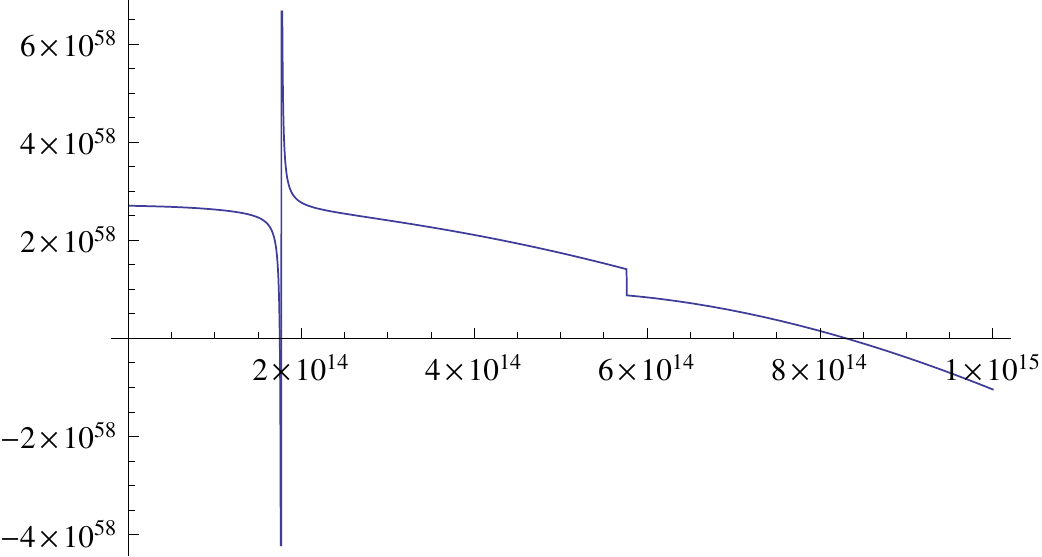}
\includegraphics[scale=0.55]{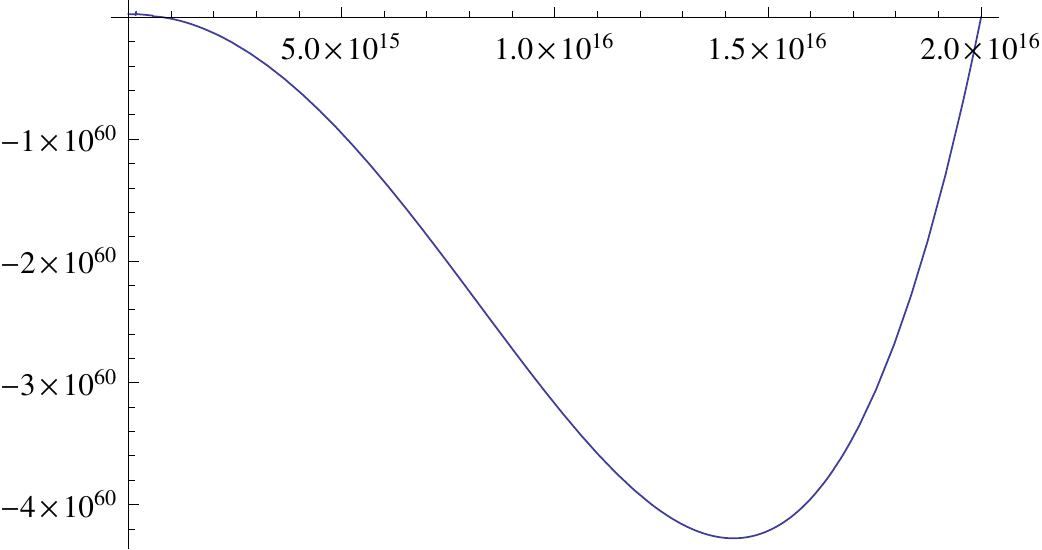}
\caption{Running of $\gamma_{0,H}(\Lambda)$ with $f_2 =1$ and with
$\gamma_0(\Lambda_{unif})=0$.\label{gamma0HFig}}
\end{figure}

To see how the interaction with the Higgs field can modify the running of the
effective cosmological constant, consider again an example that exhibits the
same behavior as the second graph of Figure \ref{gamma0Fig}. This is obtained,
for instance, by choosing $f_2=1$ and $f_4$ satisfying the vanishing condition
$\gamma_0(\Lambda_{unif})=0$. Then the running of $\gamma_{0,H}(\Lambda)$
given in \eqref{gamma0Grun} exhibits an overall behavior similar to that of
$\gamma_0(\Lambda)$ but with a change of sign and a singularity, as shown in
Figure \ref{gamma0HFig}.

\subsection{Early time bounds estimate}\label{EstimatesSec}

In models with variable cosmological (and gravitational) constant, some
strong constraints exist from ``early time bounds" on the vacuum-energy density
that are needed in order to allow nucleosynthesis and structure formation, 
see \cite{BirSar}, \cite{Freese}.

In particular, in our model, this means that we can prefer choices of the 
parameters $f_2$ and $f_4$ for which the ratio 
\begin{equation}\label{ratioeq}
\frac{\gamma_0(\Lambda)}{8 \pi G_{\rm eff}(\Lambda)} 
\end{equation}
is small at the electroweak end of the energy range we are considering, so that 
the resulting early time bound will allow the standard theory of big-bang 
nucleosynthesis to take place according to the constraints of \cite{BirSar}, \cite{Freese}.

Notice that, since we have
$\gamma_{0,H}(\Lambda)/G_{{\rm eff},H}(\Lambda)=\gamma_0(\Lambda)/G_{\rm eff}(\Lambda)$
the estimate on \eqref{ratioeq} is independent of further effects of interaction with the Higgs field,
and it only depends on the choice of the parameters $f_2$ and $f_4$ of the model. 

\begin{figure}
\includegraphics[scale=0.6]{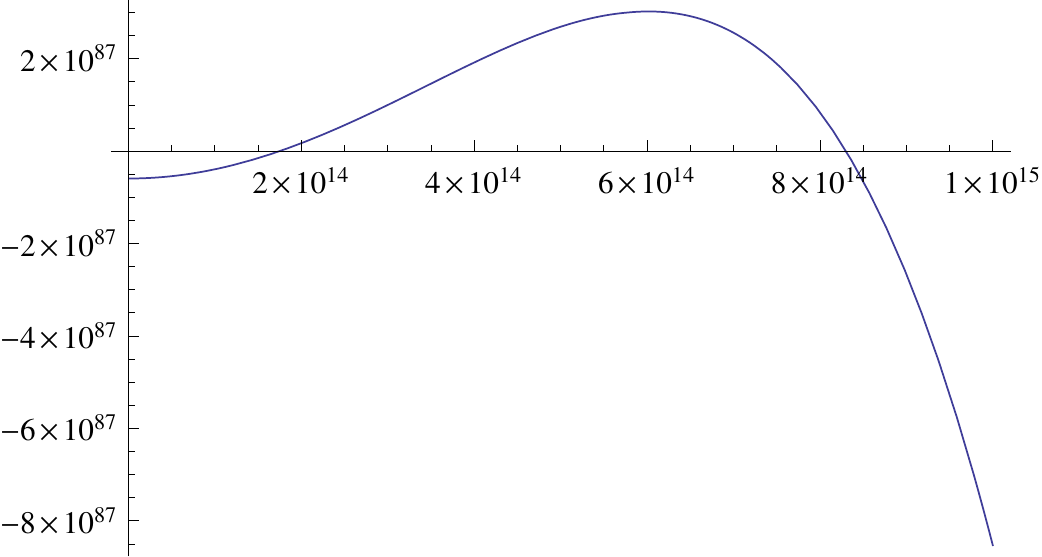}
\caption{Running of $\gamma_0(\Lambda)/(8\pi G_{\rm eff}(\Lambda))$ with $f_2 =1$ and with
$\gamma_0(\Lambda_{unif})=0$.\label{ratioFig}}
\end{figure}

This type of estimate can be used to select regions of the space of parameters $f_2$ and $f_4$
that are excluded by producing too large a value of \eqref{ratioeq}
at $\Lambda =\Lambda_{ew}$. For example, consider the two cases of Figure \ref{gamma0Fig}, 
where we set $f_2=1$ and we choose $f_4$ so that it gives vanishing $\gamma_0(\Lambda)$ 
at $\Lambda_{ew}$ or at $\Lambda_{unif}$. 
The first case gives $\frac{\gamma_0(\Lambda_{ew})}{8 \pi G_{\rm eff}(\Lambda_{ew})}=0$,
while the second gives a very large negative value of \eqref{ratioeq} at $\Lambda_{ew}$ of
$-5.93668 \times 10^{86}$ with a running of \eqref{ratioeq} as in Figure \ref{ratioFig}.

\section{Dark matter}\label{DarkSec}

We have concentrated in this paper on early universe models, with an emphasis on
various mechanisms for inflation and dark energy that arise within the NCG model,
and their mutual interactions. Another main question about cosmological implications
of the noncommutative geometry models of particle physics is whether they can accommodate
possible models for cold dark matter. This exists the domain of validity of the asymptotic expansion
of the spectral action, as such models apply to a more modern universe than what is covered by
the perturbative analysis of the spectral action functional. However, one can at least comment 
qualitatively on possible candidates within the NCG model for dark matter particles.
In addition to the minimal Standard Model, one recovers form the computation of the
asymptotic formula for the spectral action and from the fermionic part of the action in \cite{CCM}
additional right handed neutrinos with lepton mixing matrix and Majorana mass terms. Thus,
since this is at present the only additional particle content beyond the minimal standard model
that can be accommodated in the NCG setting, it is natural to try to connect this model to
existing dark matter models based on Majorana mass terms for right handed neutrinos. 

Those that seem more closely related to what one can get within the NCG model are
the ones described by Shaposhnikov--Tkachev \cite{ShapoTka}, Shaposhnikov \cite{Shapo},
and Kusenko \cite{Kusenko}. In these dark matter models, one has the usual active neutrinos,
with very small masses, and an additional number of sterile neutrinos with Majorana masses.
In the case of the $\nu$MSM model of \cite{Shapo}, one has three active and three sterile
neutrinos. In these models the sterile neutrinos provide candidate dark matter particles.
However, for them to give rise to plausible dark matter models, one needs at least one (or
more) of the sterile neutrino Majorana masses to be below the electroweak scale. In the
detailed discussion given in \cite{Kusenko} one sees that, for example, one could have two
of the the three Majorana masses that remain very large, well above the electroweak scale,
possibly close to unification scale, while a third one lowers below the electroweak scale,
so that the very large Majorana masses still account for the see-saw mechanism, while the 
smaller one provides a candidate dark matter particle.

It is possible to obtain a scenario of this kind within the NCG model, provided that one
modifies the boundary conditions of \cite{AKLRS} in such a way that, instead of having three
see-saw scales within the unification and the electroweak scale, with the smallest one already
at at very high energy around $10^{12}$ GeV, one sets things so that the lowest Majorana mass
descends below the electroweak scale. A more detailed analysis of such models will be carried
out in forthcoming work where we analyze different choices of boundary conditions for the
RGE flow of the model.

\section{Conclusions and perspectives}

We have shown in this paper how various cosmological models arise
naturally from the asymptotic expansion of the spectral action functional
in the noncommutative geometry model of particle physics of \cite{CCM}
and the running of the coefficients of this asymptotic expansion via the
renormalization group equations of \cite{AKLRS}. 

We have seen in particular the spontaneous
emergence of conformal gravity and Hoyle--Narlikar cosmologies
at phase transitions caused by the running of the effective gravitational constant.
We described effects of this running on the gravitational waves and on primordial
black holes. 
We described mechanisms by which this running, combined with the conformal
coupling of gravity to the Higgs field, can generate regions of negative gravity
in the early universe. We discussed the running of the effective cosmological constant
and slow-roll inflation models induced by the coupling of the Higgs to gravity.
We discussed briefly the connection to dark matter models based on right
handed neutrinos and Majorana mass terms. 

A planned continuation of this investigation will cover the following topics:
\begin{itemize}
\item Varying boundary conditions for the RGE flow.
\item Dark matter models based on Majorana sterile neutrinos.
\item Extensions of the NCG model with dilaton field.
\item Nonperturbative effects in the spectral action.
\item Exclusion curve from spectral index 
and tensor to scalar ratio.
\end{itemize}

\section{Appendix: boundary conditions for the RGE equations}\label{BoundarySec}

We recall here the boundary conditions for the RGE flow of \cite{AKLRS} and we
discuss the compatibility with the condition assumed in \cite{CCM}. A more
detailed analysis of the RGE flow of \cite{AKLRS} with different boundary conditions
and its effect on the gravitational and cosmological terms will be the focus of a followup 
investigation.

\subsection{The default boundary conditions}

The boundary conditions for the RGE flow equations we used in this
paper are the default boundary conditions assumed in \cite{AKLRS}.
These are as follows.

$$ \lambda(\Lambda_{unif})= \frac{1}{2} $$
$$ \mass_u(\Lambda_{unif})= \left( \begin{array}{ccc}
5.40391\times 10^{-6} & 0 & 0 \\
0 & 0.00156368 & 0 \\
0 & 0 & 0.482902 
\end{array} \right) $$
For $\mass_d(\Lambda_{unif})=(y_{ij})$ they have
$$   \begin{array}{rl}
y_{11}= & 0.0000482105 - 3.382\times 10^{-15} i \\ y_{12}= & 0.000104035 + 
 2.55017\times 10^{-7} i \\  y_{13} =& 0.0000556766 + 6.72508\times 10^{-6} i \\
y_{21}=&  0.000104035 - 2.55017\times 10^{-7} i \\ y_{22}= & 0.000509279 + 
 3.38205\times 10^{-15} i  \\ y_{23}= & 0.00066992 - 4.91159\times 10^{-8} i \\
 y_{31}= & 0.000048644 - 5.87562\times 10^{-6} i \\
 y_{32}= &  0.000585302 + 
 4.29122\times 10^{-8} i \\
 y_{33} = & 0.0159991 - 4.21364\times 10^{-20} i
\end{array}  $$
$$ \mass_e(\Lambda_{unif})=\left( \begin{array}{ccc}
2.83697\times 10^{-6} & 0 & 0 \\
0 & 0.000598755 & 0 \\
0 & 0 & 0.0101789 
\end{array} \right) $$
$$ \mass_{nu}(\Lambda_{unif})= \left( \begin{array}{ccc}
1 & 0 & 0 \\
0 & 0.5 & 0 \\
0 & 0 & 0.1 Y
\end{array} \right) $$
$$ M(\Lambda_{unif})=\left( \begin{array}{ccc}
-6.01345\times 10^{14} & 3.17771\times 10^{12} & -6.35541\times 10^{11} \\
3.17771\times 10^{12} & -1.16045\times 10^{14} & 5.99027\times 10^{12} \\
-6.35541\times 10^{11} & 5.99027\times 10^{12} & -4.6418\times 10^{12}
\end{array} \right) $$

\subsection{Constraints from NCG}

There are constraints on the boundary conditions at unification in the
noncommutative geometry model. Those were described in \cite{CCM} and we
report them here below. 
\begin{itemize}
\item A constraint on the value at unification of the parameter $\lambda$:
$$ \lambda(\Lambda_{unif})=\frac{\pi^2}{2 f_0} \frac{\fb(\Lambda_{unif})}{\fa(\Lambda_{unif})^2} $$
\item A relation between the parameter $\fa$ and the Higgs vacuum:
$$ \frac{\sqrt{\fa f_0}}{\pi} = \frac{2 M_W}{g} $$
\item A constraint on the coefficient $\fc$ at unification, coming from the see-saw
mechanism for the right handed neutrinos:
$$   \frac{2 f_2 \Lambda_{unif}^2}{f_0} \leq  \fc(\Lambda_{unif}) \leq \frac{6 f_2 \Lambda_{unif}^2}{f_0} $$
\item The mass relation at unification: 
\begin{equation}\label{massrelation}
\sum_{generations} ( m_\nu^2 + m_e^2 + 3 m_u^2 + 3 m_d^2 ) |_{\Lambda=\Lambda_{unif}} = 8 M_W^2 |_{\Lambda=\Lambda_{unif}} ,
\end{equation}
where $m_\nu$, $m_e$, $m_u$, and $m_d$ are the masses of the leptons
and quarks, that is, the eigenvectors of the matrices $\delta_{\uparrow 1}$, $\delta_{\downarrow 1}$,
$\delta_{\uparrow 3}$ and $\delta_{\downarrow 3}$, respectively, and  
$M_W$ is the W-boson mass.
\end{itemize}
Clearly, not all of these constraints are 
compatible with the default boundary conditions of \cite{AKLRS}. So either
one relaxes some of these conditions, as we have been doing in the present
paper, or one performs a wider search for more appropriate and fine tuned
boundary conditions for the RGE flow, analyzing 
how different choices of boundary values affect the behavior analyzed in this
paper. This is presently under investigation.

\end{document}